\documentclass[showpacs,preprintnumbers,amsmath,amssymb,floatfix,nofootinbib]{revtex4-1}
\usepackage[usenames]{color}
\usepackage{epsfig}
\usepackage{amssymb}
\usepackage{multirow}
\usepackage{xcolor}
\newcommand{\be}{\begin{equation}}
\newcommand{\ee}{\end{equation}}
\newcommand{\bee}{\begin{eqnarray}}
\newcommand{\eee}{\end{eqnarray}}

\renewcommand{\i}{{\mathrm{i}}}

\def \red{\color{black}}

\definecolor{grey}{rgb}{0.9,0.9,0.9}
\definecolor{black}{rgb}{0,0,0}

\def \irbaddress{Rudjer Bo\v{s}kovi\'{c} Institute, Bijeni\v{c}ka cesta 54, P.O. Box 180, 10002 Zagreb, Croatia}
\def \untzaddress{University of Tuzla, Faculty of Science, Univerzitetska 4, 75000 Tuzla, Bosnia and Herzegovina}

\begin{document}

\title{Poles of Karlsruhe-Helsinki KH80 and KA84 solutions \\  extracted by using the Laurent+Pietarinen method}

\author{Alfred \v{S}varc}
\email{alfred.svarc@irb.hr}
\affiliation{\irbaddress}
\author{Mirza Had\v{z}imehmedovi\'{c}}
\affiliation{\untzaddress}
 \author{ Rifat Omerovi\'{c}}
\affiliation{\untzaddress}
\author{Hedim Osmanovi\'{c}}
\affiliation{\untzaddress}
\author{Jugoslav Stahov}

\affiliation{\untzaddress}

\begin{abstract}

Poles of partial wave scattering matrices in hadron spectroscopy  have recently been established as a sole link 
%Comment: not sure what you mean by a sole link.
between experiment and QCD theories and models. Karlsruhe-Helsinki (KH) partial wave analyses have been ``above the line" in the Review %%@
of Particle Physics (RPP) for over three decades.
The RPP compiles Breit-Wigner (BW) parameters from local BW fits,  but give only a limited number of pole positions using speed plots %%@
(SP). 
In the KH method only Mandelstam analyticity is used as a theoretical constraint, so these partial wave solutions are as model %%@
independent as possible. 
They are a valuable source if information.
It is unsatisfactory that BW parameters given in the RPP  have been obtained from the KH80 solution, while pole parameters have been %%@
obtained from the KA84  version.
To remedy this, we have used a newly developed  Laurent+Pietarinen expansion method  to obtain pole positions for all partial waves for %%@
KH80 and KA84 solutions.
We show that differences from  pole parameters are, with a few exceptions, negligible for most partial waves.   
We give a full set of pole parameters for both solutions.

\end{abstract}


\date{\today}

\maketitle

\section{Introduction}
{\red
Revisions to the Review of Particle Physics (RPP) \cite{PDG} and contributions to recent workshops %%@
\cite{Camogli2012,Kloster2013,Camogli2013} have emphasized that poles, rather than Breit-Wigner parameters,  quantify resonance masses %%@
and widths and make a link between scattering theory and QCD. 
It appears that  Karlsruhe-Helsinki partial wave analyses make one of the most reliable data analyses ``above the line" in the RPP %%@
\cite{PDG} for almost three decades, and give Breit-Wigner parameters over local energy ranges. The pole parameters are given for only %%@
some of them. They are presently extracted from speed plots (SP) as described by  H\"{o}hler in \cite{Hoehler92,Hoehler93,Hoehler2001}.   %%@
Let $W$ be the the center of mass energy. In Ref.~\cite{Hoehler93} it was shown that by using $SP(W)=|{\rm d}T(W)/{\rm d}W|$, and  %%@
$T$-matrices defined as $T(W)=T_b + R {\Gamma} e^{ \i \, \phi}/(M-W -  \i \, {\Gamma}/2 ) $ where $T_b$ is background term,  methods of %%@
Ref.~\cite{Hoehler92} are inadequate  because the phase information is not obtained. Reference~\cite{Hoehler93}  proposed  an %%@
improvement  by introducing Argand plots for ${\rm d}T(W)/{\rm d}W$. This work assumed that ${\rm d}T_b(W)/{\rm d}W$ can be neglected. %%@
This succeeded for several partial waves, but not all.  Auxiliary assessments were recommended; 4-star resonances were derived from %%@
speed plots and Argand diagrams ${\rm d}T/{\rm d}W$  over the range $W=M \pm \Gamma /2$. The locations of $T(M)$ and $T(M \pm \Gamma %%@
/2)$ in the Argand plot for $T(W)$ were calculated by interpolating the partial wave solution KA84.  Next,  the radius  $R$ and phase %%@
$\phi$ were used to fit the resonance loop, assuming that background  $T_b$ is constant over the range $W=M \pm \Gamma /2$.  The %%@
expressions $\tilde{T}(M)$ and $\tilde{T}(M \pm \Gamma /2)$ denote   points calculated for parameters $R$ and $\phi$.  
It is shown that $T(M)$ and  $\tilde{T}(M)$  agree using this construction; however values $T(M \pm \Gamma /2)$ and $\tilde{T}(M \pm %%@
\Gamma /2)$ are  in general not yet quite satisfactory. 
Therefore $\Gamma$, $R$ and $\phi$ were adjusted until a good fit was obtained. This procedure was successful for eight of the 4-star %%@
resonances, and the parameters are listed in Table 1 of Ref. \cite{Hoehler93}.  This Table, based on  the KA84 solution, is now cited %%@
in the RPP as KH pole positions.}
\\ \\ \noindent
 In summary, the SP method is actually a three step procedure: i) make a classic speed plot; ii) make an Argand plot for ${\rm %%@
d}T(W)/{\rm d}W$ and establish a phase $\phi$; iii) correct $\Gamma$, $R$ and $\phi$ so that the interpolated value of KH (or any %%@
other) amplitude and Argand plot coincide.  We repeat the description of this procedure in detail because some younger colleagues use %%@
only the first step because H\"ohler's references \cite{Hoehler93} and \cite{Hoehler92} are not easily accessible. 
\\ \\ \noindent
This opens two issues: i) is the generalized SP method able to find all poles in KH amplitudes?  And secondly;  ii) are SP pole %%@
parameters obtained from KH80 solution comparable with  the pole position from KA84?
 So, the question arises: ``How similar these two solutions are?" Here we answer both issues. 
\\ \\ \noindent
Regarding issue i) we use the newly developed Laurent+Pietarinen method (L+P method) \cite{Svarc2012,Svarc2013} to extract all visible %%@
poles from KH80 and KA84 solutions. Differences between these solutions are quantified here. 
\\ \\ \noindent
Regarding issue ii), it is known from \cite{Hoehler93,Hoehler2001} that these two solutions are not drastically different, but are %%@
definitely not identical. 
\\ \\ \noindent
As other analyses have shown that the SP method is only a first order approximation of more general search methods %%@
\cite{Ceci2008,Masjuan2013},  it remains a mystery to us why other methods have not been used to complete the fragmentary list of KH %%@
pole parameters obtained by using SP technique only.  
\\ \\ \noindent
 So, the main purpose of this paper is to remedy these problems. We use the recent Laurent+Pietarinen (L+P) method %%@
\cite{Svarc2012,Svarc2013} to extract pole positions from both  KH80 and KA84 solutions.  We show figures and pole parameters for both, %%@
and compare them. We find more poles than originally established by the SP technique, and confirm that differences between the two sets %%@
of KH solutions are negligible. All results agree well with present results displayed in the RPP \cite{PDG}. 

\section{Formalism}
\subsection{Two classic partial wave analyses}
For almost three decades two significantly  different partial wave analyses have appeared ``above the line" in the RPP: the %%@
Carnegie-Mellon-Berkeley  (CMB) analysis of Cutkosky et al. \cite{CMB1,CMB2,CMB3}, and the Karlsruhe-Helsinki analysis by H\"{o}hler et %%@
al. \cite{Hoehler84}. 
These two analyses enforce  slightly different criteria.  The CMB model  \cite{CMB1,CMB2,CMB3} produced partial wave poles directly, %%@
but had some problems with Breit-Wigner parameters \cite{CMB3}. The Karlstuhe-Helsinki  approach was much more successful in %%@
stabilizing solutions, but had some problems in extracting BW parameters and poles.  
 
\subsubsection{CMB model}
 Ref. \cite{CMB1} amalgamated and stabilized the data base; in Ref. \cite{CMB2} they performed a single-energy stabilized partial wave %%@
analysis, and in Ref. \cite{CMB3} developed a global solution; 
a coupled-channel model with analyticity and unitarity explicitly included which they use to fit partial wave data of Ref. \cite{CMB2}. 
They explicitly get partial wave poles by analytic continuation  into the complex energy plane, but have some problems defining an %%@
``analog" to BW parameters. 
They do not make a local BW fit, but use their coupled-channel model to extract BW  parameters. 
Their list of BW parameters and poles is complete.

\subsubsection{KH80 method} The 
KH80 method used a different approach. 
Instead of performing energy stabilization at the level of partial waves, they evaluated  pion-nucleon invariant amplitudes using %%@
forward dispersion relations and the Pietarinen expansion. 
The stabilization method is close to model independent. The only constraint used is Mandelstam analyticity. 
The method includes amplitude analysis at fixed $t$, amplitude analysis at fixed cm angles, backward amplitude analysis 
and ordinary energy independent partial wave analysis, all of them linked into one computer program. 
\\ \\ \indent
The fixed$-t$ amplitude analysis used $C^{\pm}$, and $B^{\pm}$ invariant amplitudes over a large angular domain in the range $-1\le t %%@
(GeV^2)\le 0$;
invariant amplitudes satisfy  exact fixed$-t$ analyticity and $s-u$ crossing symmetry. 
Data are available up to lab. momentum $k= 200 GeV/c$. 
\\ \\ \indent
The analysis at fixed c.m. scattering  angle was done at 18 angles with $-0.8\le \cos \theta \le+0.8$..
 Forward and backward amplitude analyses were performed separately \cite{Hoehler84}.
\\ \\ \indent
Energy independent partial wave analysis is the third  step in the Karlsruhe method. 
Partial waves  were fitted to data and to invariant amplitudes at fixed$-t$, fixed c.m. scattering angle, backward and forward at the %%@
same momentum and energy. 
The strength of the constraints was adjusted allowing the possibility of weak resonances. 
Partial waves found in one iteration were used to reconstruct invariant amplitudes iteratively. 
The whole method converged in several iterations \cite{Hoehler84}. 
\\ \\ \indent
The final step in the KH method was to fit partial waves to experimental data. 
For that reason partial waves from KH80 partial wave analysis are only approximately smooth  as a function of energy. 
A smoother solution KA84 \cite{Koch85} was constructed from KH80 work using constraints from s-channel partial wave dispersion %%@
relations, fixed-s dispersion relations and information from the nearby part of the Mandelstam double spectral %%@
function~\cite{Hoehler_at_al}.
\\ \\ \indent
 Being almost model independent and consistent with Mandelstam analiticity, the Karlsruhe-Helsinki partial wave solutions are a %%@
valuable input for extraction of resonances in the $\pi N$  system.  
We use data from the original KH code preserved by one of our collaborators (Jugoslav Stahov\footnote {Jugoslav Stahov was one of the %%@
original ``KH task force" members. }) from Tuzla.

\subsection{Laurent (Mittag-Leffler) expansion}

We generalize the  Laurent expansion to the Mittag-Leffler  theorem  \cite{Svarc2013,Mittag-Leffler}, which expresses a function in %%@
terms of its first $k$ poles 
and an entire function:

\begin{eqnarray}
\label{eq:Laurent}
T(W) &=& \sum _{i=1}^{k} \frac{a_{-1}^{(i)}}{W -W_i}+B^{L}(W);    \, \, \, \, a_{-1}^{(i)}, W _i, W \in  \mathbb{C}.
\end{eqnarray}

Here, $W$ is c.m. energy, $a_{-1}^{(i)}$ and $W_ i$ are residues and pole positions for the $i$-th pole,  and $B^{L}(W )$ is a function %%@
regular in all $W  \neq W _i$. 
It is important to note that this expansion is not a representation of the unknown function $T(W)$ in the full complex energy plane, %%@
but  is restricted to the part of the complex energy plane where the expansion converges. If we choose poles as expansion points, the %%@
Laurent series converges on the open annulus around each pole.
The outer radius of the annulus extends to the position of the next singularity (such as a nearby pole ). 
Our Laurent expansion converges on a sum of circles located at the poles, and this part of the complex energy plane in principle %%@
includes the real axes. 
By fitting the expansion (\ref{eq:Laurent}) to the experimental data on the real axis, this in principle gives exact values of %%@
$s$-matrix poles.
\\ \\ \noindent
The novelty of our approach is a particular choice for the non-pole contribution $B^{L}(W )$,
based on an expansion method used by Pietarinen for $\pi N$ elastic scattering.
\\ \\ \noindent
 Before proceeding, we briefly review this method.

\subsection{Pietarinen series}

A specific type of conformal mapping technique was proposed and introduced by Ciulli \cite{Ciulli,Ciulli1} and Pietarinen %%@
\cite{Pietarinen}, and used in the Karlsruhe-Helsinki partial wave analysis \cite{Hoehler84} as an efficient expansion of 
invariant amplitudes.  
It was later used by a  number of authors to solve problems in scattering and field theory \cite{ConMap-Use}, but not applied to the %%@
pole  search prior to our recent study~\cite{Svarc2013}. 
A more detailed discussion of the use of conformal mapping and this method can be found in Refs.\cite{Svarc2013,Mittag-Leffler}.

If $F(W)$ is a general unknown analytic function with a cut starting at $W=x_P$, it can be represented as a power series of %%@
``Pietarinen functions" 
\begin{eqnarray}
\label{eq:Pietarinen}
F(W ) &=& \sum_{n=0}^{N}c_n\, X(W )^n, \, \, \, \, \, \, \, \, \, \, W  \in  \mathbb{C}   \nonumber \\
X(W )&=& \frac{\alpha-\sqrt{x_P-W }}{\alpha+\sqrt{x_P-W }}, \, \, \, \, \, c_n, x_P, \alpha \in  \mathbb{R},
\end{eqnarray}
with $\alpha$ and $c_n$ acting as  tuning parameter and coefficients of the Pietarinen function $X(W)$ respectively.

The essence of the approach is that $(X(W )^n, \, n=1, \, \infty)$ forms a complete set of functions defined on the unit circle in the %%@
complex energy plane with a  branch cut starting at $W= x_P$; 
the analytic form of the function is initially undefined. 
The final form of the analytic function $F(W)$ is obtained by introducing a rapidly convergent power series with real coefficients, and %%@
the degree of the expansion is automatically determined by fitting the input data. 
In the calculation of  Ref.~\cite{Pietarinen}, as many as 50 terms were used; in the present analysis, covering a narrower energy %%@
range, fewer terms are required.

\subsection{Application of Pietarinen series to scattering theory}

The analytic structure of each partial wave is well known. 
Every partial wave contains poles which parameterize resonant contributions, cuts in the physical region starting at thresholds of %%@
elastic and all possible inelastic channels, plus
$t$-channel, $u$-channel and nucleon exchange contributions quantified with corresponding negative energy cuts. 
However, the explicit analytic form of each cut contribution is not known. 
Instead of guessing the exact analytic form of all of these, we use one Pietarinen series to represent each cut, and the number of %%@
terms in the Pietarinen series is determined by the quality of fit to the input data. 
In principle we have one Pietarinen series per cut; branch points $x_P, x_Q ...$ are known from physics, and coefficients are %%@
determined by fitting the input data.  
In practice, we have too many cuts (especially in the negative energy range), so we reduce their number by dividing them into two %%@
categories: 
all negative energy cuts are approximated with only one, effective negative energy cut represented by one (Pietarinen) series (we %%@
denote its branch point as $x_P$), while each physical cut is represented by a separate
 series with branch points determined by the physics of the process ($x_Q,x_R...$).

In summary, the set of equations which define the Laurent expansion + Pietarinen series method (L+P method) is:

\begin{eqnarray}
\label{eq:Laurent-Pietarinen}
T(W ) &=& \sum _{i=1}^{k} \frac{a_{-1}^{(i)}}{W-W_i}+ B^{L}(W) 
\nonumber \\
B^{L}(W)&=& \sum _{n=0}^{M}c_n\, X(W )^n  +  \sum _{n=0}^{N}d_n\, Y(W )^n +  \sum _{n=0}^{N}e_n\, Z(W )^n  + \cdots    
\nonumber  \\
X(W )&=& \frac{\alpha-\sqrt{x_P-W}}{\alpha+\sqrt{x_P - W }}; \, \, \, \, \,   Y(W ) =  \frac{\beta-\sqrt{x_Q-W }}{\beta+\sqrt{x_Q-W }};  %%@
\, \, \, \, \,   Z(W ) =  \frac{\gamma-\sqrt{x_R-W }}{\gamma+\sqrt{x_R-W }} + \cdots  
\nonumber \\
&& a_{-1}^{(i)}, W _i, W   \in   \mathbb{C} 
\nonumber \\
&& c_n, d_n, e_n \alpha, \beta, \gamma ... \in  \mathbb{R}  \, \, 
{\rm and} \, \, x_P, x_Q, x_R  \in \mathbb{R} \, \,  {\rm  or} \, \, \mathbb{C}   
\nonumber \\
&& {\rm and} \, \, \, k, M, N  ... \in  \mathbb{N}.
\end{eqnarray}

As our input data are on the real axes, the fit is performed only on this dense subset of the complex energy plane. 
All Pietarinen parameters in equations (\ref{eq:Laurent-Pietarinen}) are determined by the fit.

We observe that the class of input functions which may be analyzed with this method is quite wide.
One may either fit partial wave amplitudes obtained from theoretical models, or possibly experimental data directly. 
In either case, the $T$-matrix is represented by this set of equations (\ref{eq:Laurent-Pietarinen}), and minimization is
 usually carried out in terms of $\chi^2$.

\subsection{Real and complex branch points}
Branch points $x_P$, $x_Q$, $x_R$ ....  in the Pietarinen expansion (\ref{eq:Laurent-Pietarinen})  can be real or complex.
However, real or complex  branch points  describe different physical situation. 
If the branch points $x_P$, $x_Q$, $x_R$ .... are real numbers, this means that our background contributions are defined by stable %%@
initial and final state particles. 
Then all contributions to the observed processes are created by intermediate isobar resonances, and all other initial and final state %%@
contributions are given by stable particles,  as described by Pietarinen expansions with real branch point coefficients.  
From experience we know that this \underline{in principle} is not true: a three body final state is always created provided that the %%@
energy balance allows for it, and in three body final states we typically do have a contribution from one stable particle (nucleon or %%@
pion), and many other combinations of two-body resonant substates like $\sigma$, $\rho$, $\Delta$... {\red 
So we choose the model where the first two branch points $x_P$ and  $x_Q$ are always real, but the third branch point $x_R$ can be %%@
either real (two body final states) or complex (three body final state with a resonance in a two body subsystem).  }
\\ \\ \indent
Let us claim the fact that single channel character of the method prohibits us to establish with certainty which mechanism prevails. %%@
Using data from a single channel only (the existing KH80 and KA84 input)  we are unable with certainty to say whether the new resonant %%@
state which appears is an isobar state with two body final states, or a three body final state with a resonance in a two body %%@
subsystem. 
If only single channel information is available, we have two alternatives: either we obtain a good fit with an extra resonance and %%@
stable initial and final state particles (real branch points), or we obtain a good fit with one resonance less, and a complex branch %%@
point. 
Data from a single channel data does not distinguish between the two. 
This effect has been already spotted, elaborated and discussed in the case of J\"{u}lich model, and a more detailed elaboration how the %%@
$\rho$N complex branch point interferes and intermixes with $P_{11}$(1710) 1/2$^+$  \cite{Ceci2011,Roenchen2013}.
\\ \\ \indent
Issues connected with importance of inelastic channels, and two-body resonant sub-states in three-body final states have already been %%@
recognized in Ref. \cite{Hoehler93} (paragraphs 4.2 and 4.3).  
However, at that time, a formalism to follow and quantify these effects  didn't exist, so no estimates have been given. 
L+P formalism with complex branch points enables us to study these effects in detail.

\begin{enumerate}
\item In either case, a new resonant state is established, but our single-channel method cannot say where (either in two body %%@
intermediate state or in three-body subchannel). 
We cannot distinguish whether the new resonant state manifests itself as a new isobar resonance with stable initial and final states %%@
(real branch points), or as a resonance in two-body subchannel of three-body final state (complex branch point). 
For that, we need the data from extra channels, and an experiment giving us missing information on ratio of 2-body/3-body cross %%@
sections at the same energies is badly missing.

The advantage of the Pietarinen expansion method is that it can be extended directly to complex branch points; 
we using it to search for suspicious partial waves.
%Query: is this what you meant?
 
\item We claim that this effect is not  affecting only $P_{11}$(1710) resonance as established by the J\"{u}lich group %%@
\cite{Ceci2011,Roenchen2013} for the $\rho N$ branch point, but influences interpretation of many more resonances from the RPP (at %%@
least we have established that for the Karlsruhe-Helsinki PWA). 
One definitely  needs measurements from other channels before claiming whether the observed structure is an intermediate isobar %%@
resonance, or a resonance appearing in the two-body subsystem of a three body final state. 
Single channel measurements are insufficient, we need multi-channel measurements in order to distinguish between the two. 
H\"{o}hler has in his Newsletter's  paper \cite{Hoehler93} discussed similar problems, but he blamed  guilt on the $\omega$N branch %%@
point. 
However, in this paper we claim the effects of the $\rho$N branch point are much more pronounced. 
We have tested the influence of the better known $\pi \Delta$ branch point located at $(1370-\i \, 40)$ MeV on KH amplitudes, but as it %%@
is much lower in mass than the $\rho N$ branch point,  its influence was negligible.  

\end{enumerate}
\subsection{Fitting procedure}

We use three Pietarinen functions (one with a branch point in the unphysical region to represent all left-hand cuts, and two with  %%@
branch points in the physical region to represent the dominant inelastic channels), combined with the minimal number of poles. 
We also allow the possibility  that one of the branch points becomes a complex number allowing all three-body final states to be %%@
effectively taken into account. 
We generally start with 5 Pietarinen terms per decomposition, and the anticipated number of poles. 
The discrepancy criteria are defined below using a discrepancy parameter $D_{dp}$. 
This quantity is minimized using MINUIT and the quality of the fit is visually inspected by comparing fitting function with  data. 
If the fit is unsatisfactory (discrepancy parameters are too high, or fit visually  does not reproduce the fitted data), the number of %%@
Pietarinen terms is increased, and if it does not help, the number of poles is increased by one. 
The fit is repeated, and the quality of the fit is re-estimated. 
This procedure is continued until we reach a satisfactory fit.

Pole positions, residues, and Pietarinen coefficients $\alpha$, $\beta$, $\gamma$, $c_i$, $d_i$ and $e_i$ are 
our fitting parameters. 
However, in the strict spirit of the method, Pietarinen branch points $x_P$, $x_Q$ and $x_R$ should not be fitting parameters; 
each known cut should be represented by its own Pietarinen series, fixed to known physical branch points. 
While this would be ideal, in practice the application is somewhat different.  
We can never include all physical cuts from the multi-channel process. 

Instead, we represent them by a smaller subset. 
So, in our  method, Pietarinen branch points $x_P$, $x_Q$ and $x_R$ are not generally constants;  we have explored the effect of %%@
allowing them to vary as fitting parameters. 
In the following, we shall demonstrate that when searched, the branch points in the physical region still naturally converge towards %%@
branch points which belong to channels which dominate a particular partial wave, but may not actually correspond to them exactly.
The proximity of the fit results to  exact physical branch points describes the goodness of fit; it tells us how  well certain %%@
combinations of thresholds is indeed approximates a partial wave.
Together with the choice of the degree of Pietarinen polynomial, this represents the model dependence of our method. 
We do not claim that our method is entirely model independent. However, the method chooses the simplest function with the given %%@
analytic properties which fit the data, and increases the complexity of the function only when the data require it.  
\\ \\ 
\noindent
\subsection{Error analysis}

When we fit KH80 and KA84  amplitudes, we have to define which parameters we are minimizing. 
\\ \\ \noindent
For   both solutions we introduce the discrepancy parameter per data point $D_{dp}$ (the substitute for $\chi^2_{dp}$ per data point %%@
when analyzing experimental data):
\begin{eqnarray} \label{def:D} 
D_{dp}  &=&  \frac{1}{2 \, N_{data}} \, \, \sum_{i=1}^{N_{data}} \left[ \left( \frac{{\rm Re} T_i^{fit}-{\rm Re} T_i^{KH}}{Err_i^{\rm %%@
Re}}  \right)^2 + \right. 
\nonumber \\
 & & \hspace*{2.2cm}  \left. +   \left( \frac{{\rm Im} T_i^{fit}-{\rm Im} T_i^{KH}}{Err_i^{\rm Im}} \right)^2 \right] ,
\end{eqnarray}

where $N_{data}$ is the number of energies, and errors of   KH80 and KA84 solutions are introduced as:

\begin{eqnarray}  \label{def:Err}
Err_i^{\rm Re} & = &  0.05 \, \, \frac{ \sum_{k=1}^{N_{data}} |{\rm Re} T_k^{KH}|} {N_{data}} + 0.05 \, \, |{\rm Re} T_i^{KH}| 
\nonumber \\
Err_i^{\rm Im} & = &  0.05 \, \, \frac{ \sum_{k=1}^{N_{data}} |{\rm Im} T_k^{KH}|} {N_{data}} + 0.05 \, \, |{\rm Im} T_i^{KH}| . 
\end{eqnarray}

{\red
When errors of the input numbers are not given, and one wants to make a minimization, errors have to be estimated. 
There are two simple ways to do it:  either assigning a constant error to each data point, or introducing an energy dependent error as %%@
a percentage of the given value. 
Both definitions have drawbacks. 
For the first recipe only high-valued points are favored, while in the latter case low-valued points tend to be almost exactly %%@
reproduced. 
We find neither satisfactory, so we follow prescriptions used by GWU and Mainz groups, and use a combined error which consists of a sum %%@
of constant and energy dependent errors.} 
\\ \\ \noindent
In our principal paper \cite{Svarc2013} we have tested the validity of the model on a number of well known $\pi$N amplitudes, and %%@
concluded that the method is very robust. 
That paper did not present an error analysis. 
That is done here.
\\ \\ \noindent
In the L+P method we have statistical and systematic uncertainties: 1. statistical; and 2. systematic.
\\ \\
\noindent
\textit{1. Statistical uncertainty} 
\\ \\
Statistical uncertainties are simply taken  from MINUIT, which is used for minimization. It is shown separately in all tables as the %%@
first term. 
\\ \\ \noindent
\textit{2. Systematic uncertainty}
\\ \\
Systematic uncertainty is the error of the method itself, and requires a more detailed explanations. 
\\ \\
Our 
Laurent decomposition contains only two branch points in the physical region, and this is far from enough
 in a realistic case. 
Any realistic analytic function in principle contains more than two branch points approximated in our model by a different analytic %%@
function containing only two. 
 \\ \\ \noindent
 We use the following procedure to define systematic uncertainties:

\begin{itemize}
               
\item [i)]We release the first (unphysical) branch point $x_P$ because we have no control over background contributions;
\item [ii)] We always keep the first physical branch point $x_Q$ fixed at $x_Q=1077$ MeV (the $\pi$N threshold).
\item [iii)] The error analysis is done by varying the remaining physical branch point $x_R$ in two ways:
			       
 \begin{enumerate}
					     
\item We fix the third branch point $x_R$ to the threshold of the dominant inelastic channel for the chosen partial wave (e.g. the %%@
$\eta$ threshold for S-wave) if only one inelastic channel is important, or in case of several equally important inelastic processes we %%@
perform several runs with  the $x_R$  branch point fixed to each threshold in succession.
\item  We release the third branch point  $x_R$  allowing MINUIT to find an effective branch point representing all inelastic channels. 
If only one channel is dominant, the result of the fit will be close to the dominant inelastic channel.
                    
\end{enumerate}
					
\item [iv)] We average results of the fit, and obtain the standard deviation.
\end{itemize}

The choice of all values for the branch point $x_R$ is given in  the Appendix (Tables \ref{tab:paramKH80} for KH80 solution  and Table %%@
\ref{tab:paramKA84} for KA84). 
The quality of our fits for  both KH80 and KA84 solutions  are  measured by the discrepancy parameter $D_{dp}$ defined in Eqs. %%@
(\ref{def:D}) and (\ref{def:Err}).

\section{Results}

\subsection{Real branch points}

\begin{figure*}[!p]
\includegraphics[width=0.80\textwidth]{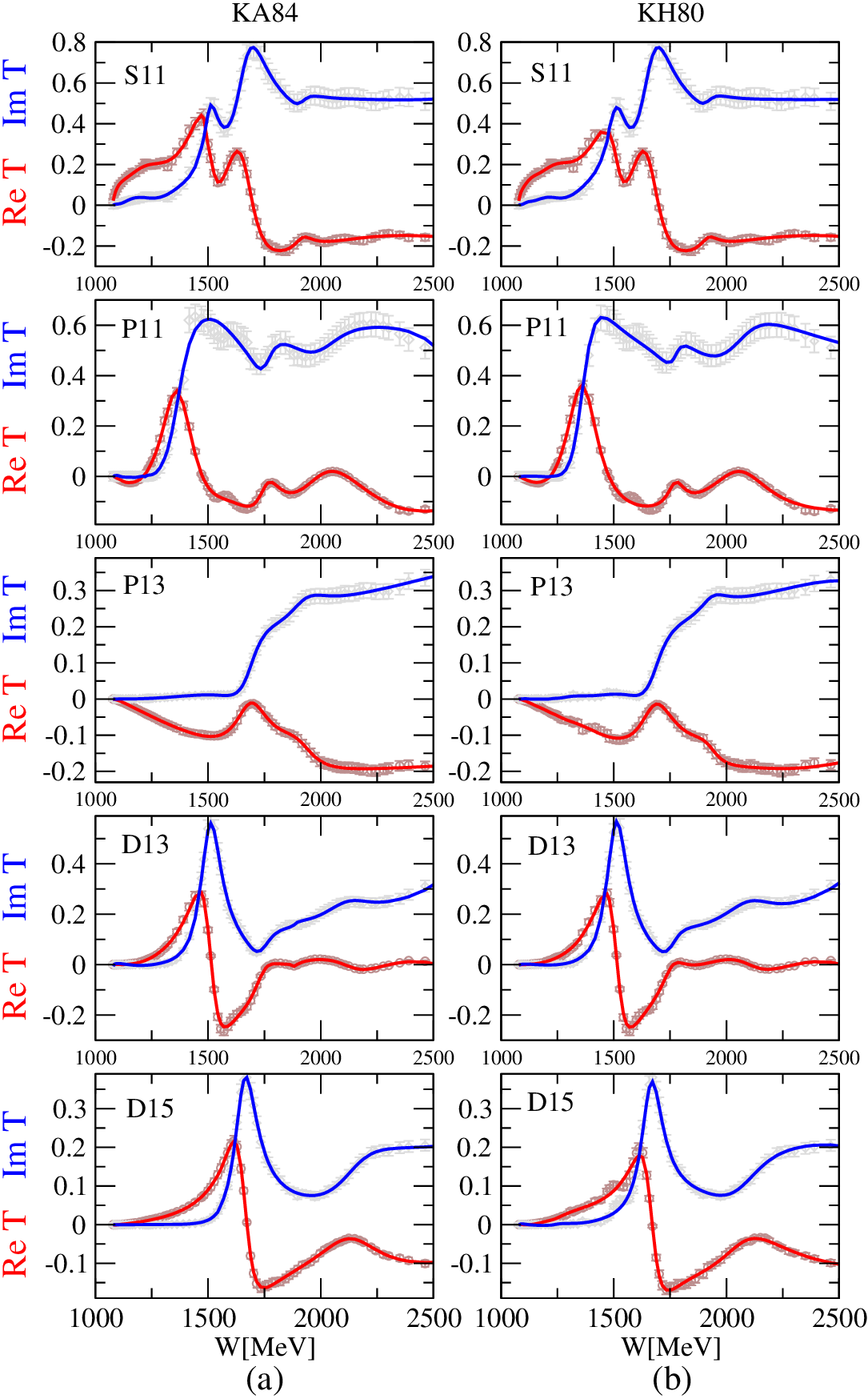}
\caption{(Color online)  L+P fit for I=1/2 solutions, (a) Fit to KA84, (b) Fit to KH80.}
\label{Fig1}
\end{figure*}

\begin{figure*}[!p]
\includegraphics[width=0.85\textwidth]{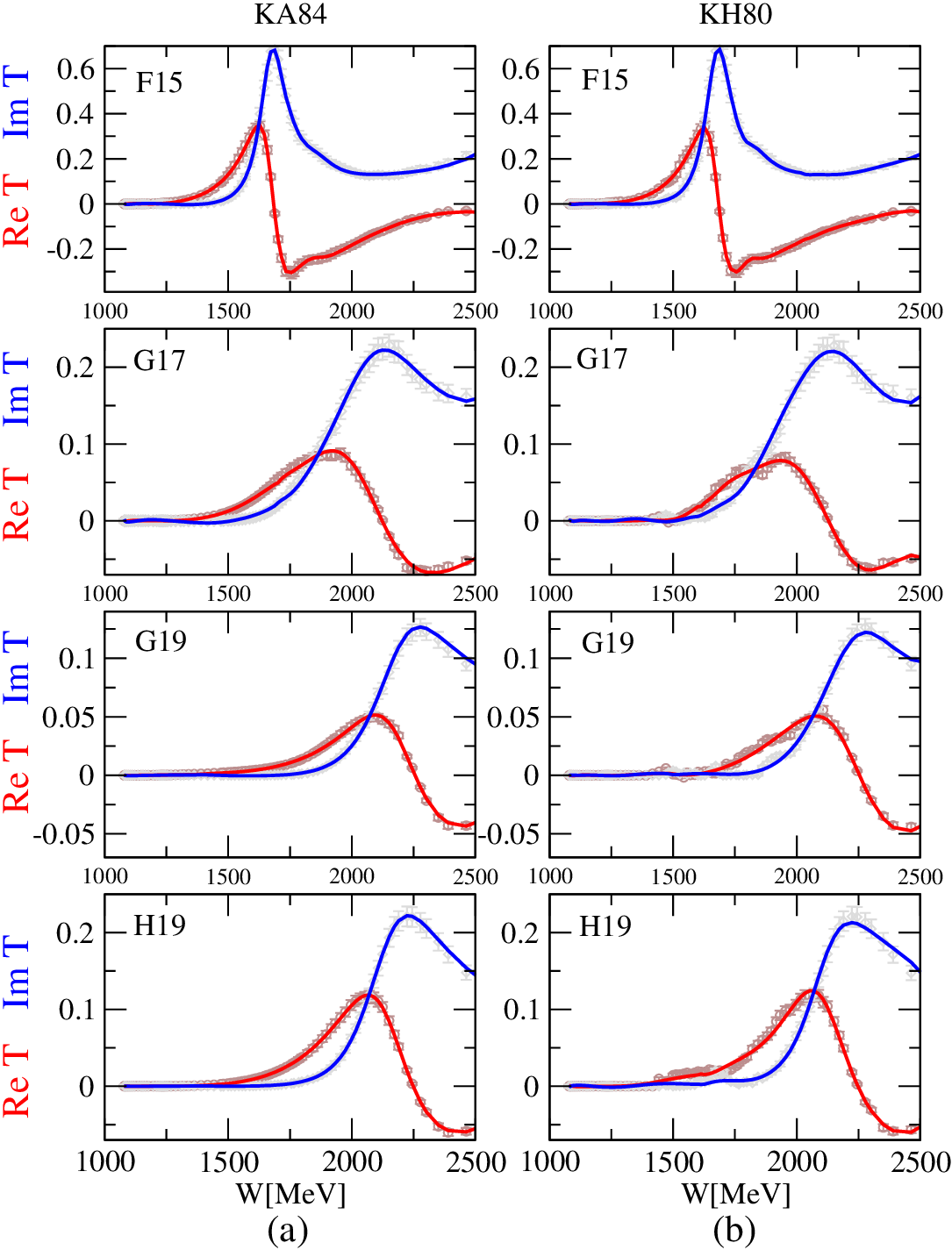}
\caption{(Color online)  L+P fit for I=1/2 solutions, (a) Fit to KA84, (b) Fit to KH80.}
\label{Fig2}
\end{figure*}

\begin{figure*}[!p]
\includegraphics[width=0.80\textwidth]{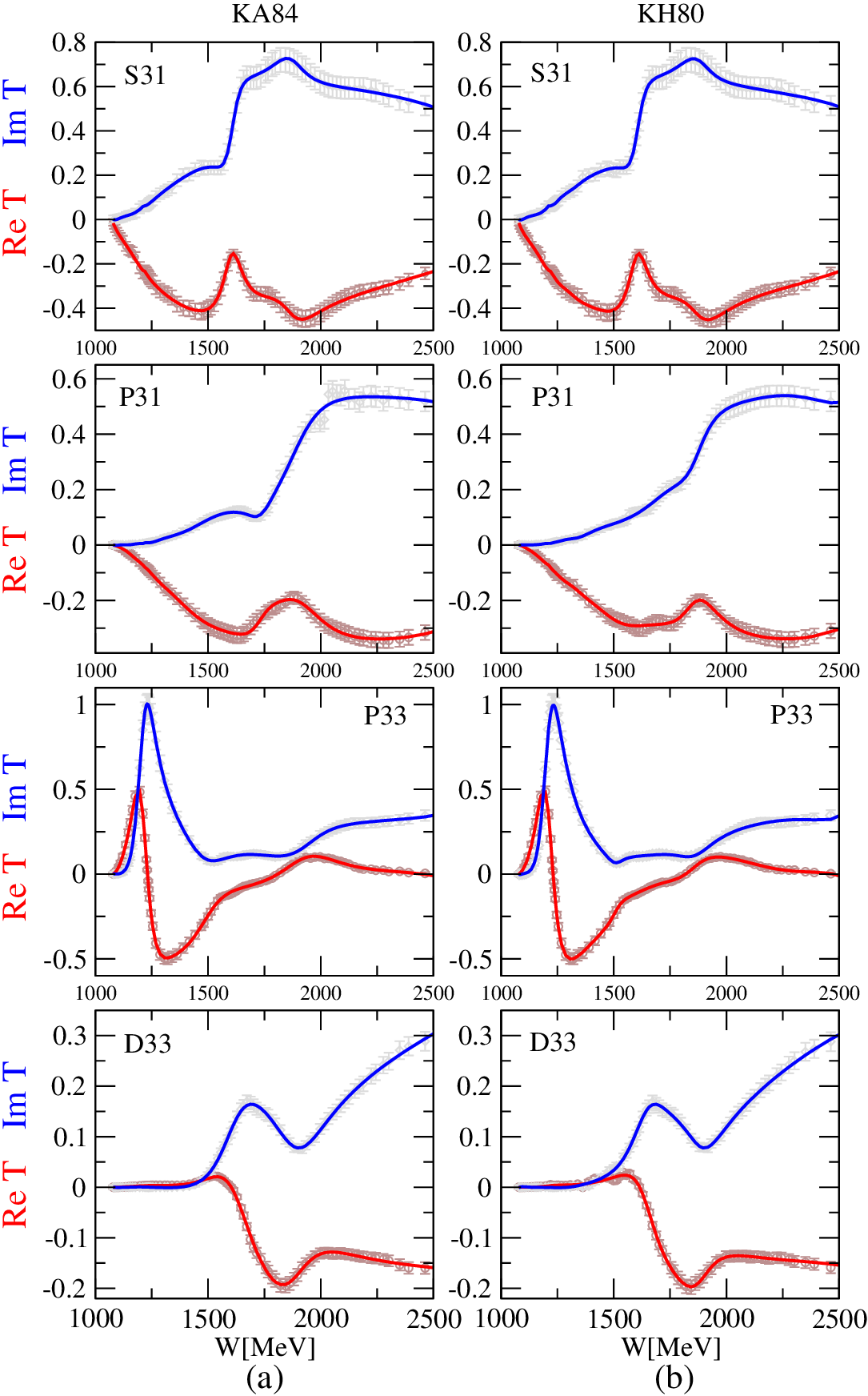}
\caption{(Color online)  L+P fit for I=3/2 solutions, (a) Fit to KA84, (b) Fit to KH80.}
\label{Fig3}
\end{figure*}

\begin{figure*}[!p]
\includegraphics[width=0.80\textwidth]{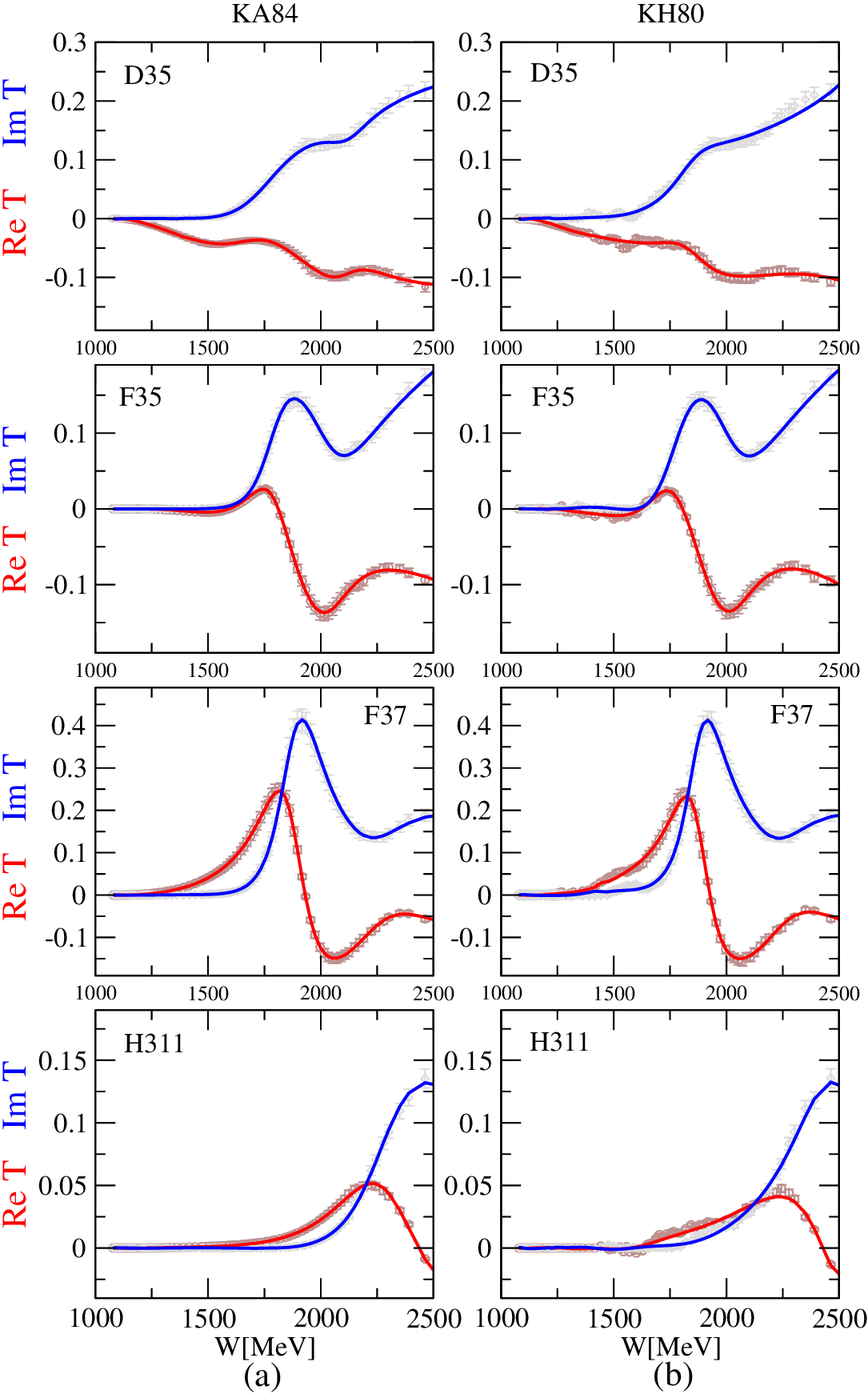}
\caption{(Color online)  L+P fit for I=3/2 solutions, (a) Fit to KA84, (b) Fit to KH80.}
\label{Fig4}
\end{figure*}

In Tables \ref{tab:pole1} -  \ref{tab:pole4} and in Figs. \ref{Fig1} - \ref{Fig4} we show L+P pole parameters and quality of the fit  %%@
for all KH80 and KA84  partial waves for the case where the  case where the
the reaction is 2-body $\rightarrow$ 2-body with an unknown number of resonances in intermediate isobar states. 
In Tables~\ref{tab:paramKH80} and \ref{tab:paramKA84}, given in the Appendix, we show corresponding L+P parameters. 
In this case three-body final states are neglected.

\begin{table*}[!hb]
 \caption{{\footnotesize \label{tab:pole1} Pole positions in MeV and residues
of partial waves as moduli in MeV and phases in degrees for lowest $I=1/2$ partial waves. 
The results 
from L+P expansion are given for Karlsruhe-Helsinki 80 (KH80) and
Karlsruhe 84 (KA84) analysis. 
Resonances marked with a star indicate resonances which can be explained by the $\rho$N complex branch point. 
{\red RPP denotes the range of pole parameters given by Ref.~\cite{PDG}, and RPP H93 denotes the values of pole parameters 
named HOEHLER 93 in the RPP, and taken from RPP and Table 1 of Ref. \cite{Hoehler93}.} 
\\ }}
\begin{tabular}{cc|ccccc}
\hline 
PW  & Source  & \textcolor{black}{Resonance}  & \textcolor{black}{$\;\mbox{Re}\, W_{p}\;$}  & \textcolor{black}{$\;-2\mbox{Im}\, %%@
W_{p}\;$}  & \textcolor{black}{$|\mbox{residue}|$}  & \textcolor{black}{$\theta$} \tabularnewline
\hline 
\hline 
\multirow{12}{*}{ $S_{11}$ } 
& RPP  &  & \textbf{$\mathbf{1490-1530}$} & \textbf{$\mathbf{90-250}$} & \textbf{$\mathbf{50\pm20}$} & %%@
\textbf{$\mathbf{(-15\pm15)^{\circ}}$}\tabularnewline
 
& RPP H93  &  & \textbf{$\mathbf{1487}$} & \textbf{$-$} & \textbf{$-$} & \textbf{$-$}\tabularnewline
 & \multicolumn{1}{c|}
{KH80 L+P } & \textbf{$N(1535) \:1/2^{-}$}  & $1509\pm4\pm2$ & $118\pm9\pm2$ & $22\pm2\pm0.4$ & $(-5\pm5\pm3)^{\circ}
$\tabularnewline
 & KA84 L+P & & $1505\pm3\pm1$ & $103\pm7\pm3$ & $20\pm2\pm1$ & $(-14\pm3\pm1)^{\circ}
$\tabularnewline
\cline{2-7} 
 
& RPP  &  & $\mathbf{1640-1670}$ & $\mathbf{100-175}$ & $\mathbf{20-50}$ & \textbf{$\mathbf{(-50-80)^{\circ}}$}
\tabularnewline
 & RPP H93 &  & $\mathbf{1670}$ & $\mathbf{163}$ & $\mathbf{39}$ & \textbf{$\mathbf{-37^{\circ}}$}
\tabularnewline
 & KH80 L+P  & $N(1650)\:1/2^{-}$ & $1660\pm3.5\pm1$ & $167\pm8\pm2$ & $47\pm3\pm1$ & $(-47\pm3\pm1)^{\circ}
$\tabularnewline
 & KA84 L+P &  & $1663\pm3\pm0$ & $165\pm7\pm1$ & $45\pm2\pm1$ & $(-44\pm3\pm1)^{\circ}
$\tabularnewline
\cline{2-7} 
 
& RPP  &  & \textbf{$\mathbf{1900-2150}$} & \textbf{$\mathbf{90-479}$} & \textbf{$\mathbf{1-60}$} & $\mathbf{(0-164)^{\circ}}
$\tabularnewline
 & RPP H93  &  & \textbf{$-$} & \textbf{$-$} & \textbf{$-$} & $-$\tabularnewline
 & KH80 L+P  & $N(1895)\:1/2^{-}$ & $1917\pm19\pm1$ & $101\pm36\pm1$ & $3.1\pm1.4\pm0$ & $(-107\pm23\pm2)^{\circ}$\tabularnewline
 & KA84 L+P &  & $1920\pm19\pm2$ & $93\pm15\pm3$ & $2.7\pm1\pm0.2$ & $(-105\pm23\pm3)^{\circ}$\tabularnewline
\hline\hline
\multirow{12}{*}{$P_{11}$ } & RPP &  & $\mathbf{1350-1380}$ & $\mathbf{160-220}$ & $\mathbf{40-52}$ & %%@
$\mathbf{(-75-100)^{\circ}}$\tabularnewline
 & RPP H93 &  & $\mathbf{1385}$ & $\mathbf{164}$ & $\mathbf{40}$ & $-$\tabularnewline
 & KH80 L+P  & $N(1440)\:1/2^{+}$ & $1363\pm2\pm2$ & $180\pm4\pm5$ & $50\pm1\pm2$ & $(-88\pm1\pm2)^{\circ}$\tabularnewline
 & KA84 L+P &  & $1365\pm2\pm4$ & $187\pm4\pm10$ & $48\pm1\pm3$ & $(-88\pm1\pm4)^{\circ}$\tabularnewline
\cline{2-7} 
 & RPP & & $\mathbf{1670-1770}$ & $\mathbf{80-380}$ & $\mathbf{6-15}$ & %%@
${\color{black}{\color{black}\mathbf{(90-200)^{\circ}}}}$\tabularnewline
 & RPP H93  &  & $\mathbf{1690}$ & $\mathbf{200}$ & $\mathbf{15}$ & $-$\tabularnewline
 
& KH80 L+P  & $N(1710)^*\:1/2^{+}$   & $1770\pm5\pm2$ & $98\pm8\pm5$ & $5\pm1\pm1$ & $(-104\pm7\pm3)^{\circ}
$\tabularnewline
 & KA84 L+P &  & $1763\pm4\pm9$ & $105\pm5\pm10$ & $6\pm1\pm1$ & $(-117\pm4\pm15)^{\circ}
$\tabularnewline
\cline{2-7} 
 
& RPP  & & $\mathbf{2120\pm40}$ & $\mathbf{180-420}$ & $\mathbf{14\pm7}$ & $\mathbf{(35\pm25)^{\circ}}$\tabularnewline
 
& RPP H93 &  & - & - & - & -
\tabularnewline
 & KH80 L+P  & $N(2100)^*\:1/2^{+}$  & $2052\pm6\pm3$ & $337\pm10\pm4$ & $30\pm1\pm1$ & $(-92\pm3\pm2)^{\circ}
$\tabularnewline
 & KA84 L+P &  & $2023\pm5\pm25$ & $346\pm9\pm13$ & $32\pm1\pm3$ & $(-118\pm3\pm21)^{\circ}
$\tabularnewline
\hline \hline
\multirow{8}{*}{$P_{13}$} & RPP  & & $\mathbf{1660-1690}$ & $\mathbf{150-400}$ & $\mathbf{15\pm8}$ & $\mathbf{(-130\pm30)^{\circ}}
$\tabularnewline
 & RPP H93 &  & $\mathbf{1686}$ & $\mathbf{187}$ & $\mathbf{15}$ & $-$
\tabularnewline
 & KH80 L+P  & $N(1720)\:3/2^{+}$  & $1677\pm4\pm1$ & $184\pm8\pm1$ & $13\pm1\pm0$ & $(-115\pm3\pm2)^{\circ}
$\tabularnewline
 & KA84 L+P &  & $1685\pm4\pm1$ & $178\pm8\pm1$ & $13\pm1\pm1$ & $(-104\pm4\pm1)^{\circ}
$\tabularnewline
\cline{2-7} 
 
& RPP  & & $\mathbf{1870-1930}$ & $\mathbf{140-300}$ & $\mathbf{3\pm2}$ & $\mathbf{(10\pm35)^{\circ}}
$\tabularnewline
 & RPP H93 &  & $-$ & $-$ & $-$ & $-$
\tabularnewline
 & KH80 L+P  & $N(1900)^*\:3/2^{+}$  & $1928\pm18\pm2$ & $152\pm40\pm9$ & $4\pm1\pm1$ & $(-29\pm15\pm2)^{\circ}$
\tabularnewline
 & KA84 L+P &  & $1920\pm17\pm1$ & $215\pm37\pm2$ & $7\pm1\pm1$ & $(-38\pm11\pm1)^{\circ}
$\tabularnewline
\hline \hline
\end{tabular}
\end{table*}

\begin{table*}[!p]
 \caption{{\footnotesize \label{tab:pole2} Pole positions in MeV and residues
of partial waves as moduli in MeV and phases in degrees for higher $I=1/2$ partial waves. The results
from L+P expansion are given for Karlsruhe-Helsinki 80 (KH80) and
Karlsruhe 84 (KA84) analysis. Resonances marked with a star indicate resonances which can be explained by $\rho$N complex branch point. %%@
{\red RPP denotes the range of pole parameters given by Ref.~\cite{PDG}, and RPP H93 denotes the values of pole parameters named %%@
HOEHLER 93 in RPP, and taken over from RPP and Table 1 of Ref. \cite{Hoehler93}.} \\ }}
\begin{tabular}{cc|ccccc}
\hline \hline
PW  & Source  & \textcolor{black}{Resonance}  & \textcolor{black}{$\;\mbox{Re}\, W_{p}\;$}  & \textcolor{black}{$\;-2\mbox{Im}\, %%@
W_{p}\;$}  & \textcolor{black}{$|\mbox{residue}|$}  & \textcolor{black}{$\theta$} \tabularnewline
\hline 
\hline 
\multirow{12}{*}{$D_{13}$} & RPP  &  & $\mathbf{1505-1515}$ & $\mathbf{105-120}$ & $\mathbf{35\pm3}$ & %%@
$\mathbf{(-10\pm5)^{\circ}}$\tabularnewline
 & RPP H93 &  & $\mathbf{1510}$ & $\mathbf{120}$ & $\mathbf{32}$ & $\mathbf{-8^{\circ}}$\tabularnewline
 & KH80 L+P  &  \textcolor{black}{$N(1520)\;3/2^{-}$} & $1506\pm1\pm1$ & $115\pm2\pm1$ & $33\pm1\pm1$ & %%@
$(-15\pm1\pm1)^{\circ}$\tabularnewline
 & KA84 L+P &  & $1506\pm1\pm1$ & $116\pm2\pm2$ & $33\pm1\pm1$ & $(-15\pm1\pm1)^{\circ}$\tabularnewline
\cline{2-7} 
 & RPP  &  & $\mathbf{1650-1750}$ & $\mathbf{100-350}$ & $\mathbf{5-50}$ & $\mathbf{(-120\: to\:20)^{\circ}}$\tabularnewline
 & RPP H93 &  & $\mathbf{1700}$ & $\mathbf{120}$ & $\mathbf{5}$ & $-$\tabularnewline
 & KH80 L+P  & $N(1700)^*\:3/2^{-}$ & $1757\pm4\pm1$ & $136\pm7\pm4$ & $7\pm1\pm1$ & $(-113\pm4\pm2)^{\circ}$\tabularnewline
 & KA84 L+P &  & $1743\pm4\pm4$ & $132\pm7\pm2$ & $7\pm1\pm1$ & $(-134\pm4\pm6)^{\circ}$\tabularnewline
\cline{2-7} 
 & RPP  & & $\mathbf{1800-1950}$ & $\mathbf{150-250}$ & $\mathbf{2-10}$ & $\mathbf{(180\pm80)^{\circ}}$\tabularnewline
 & RPP H93 &  & $-$ & $-$ & $-$ & $-$\tabularnewline
 & KH80 L+P  & $N(1875)^*\:3/2^{-}$  & $2094\pm7\pm11$ & $296\pm15\pm4$ & $13\pm1\pm1$ & $(-2\pm4\pm9)^{\circ}$\tabularnewline
 & KA84 L+P &  & $2120\pm6\pm11$ & $270\pm13\pm5$ & $11\pm1\pm1$ & $(17\pm4\pm5)^{\circ}$\tabularnewline
\hline \hline
\multirow{8}{*}{$D_{15}$} & RPP  &  & $\mathbf{1655-1665}$ & $\mathbf{125-150}$ & $\mathbf{25\pm5}$ & %%@
$\mathbf{(-25\pm6){}^{\circ}}$\tabularnewline
 & RPP H93  &  & $\mathbf{1656}$ & $\mathbf{126}$ & $\mathbf{23}$ & $\mathbf{-22^{\circ}}$\tabularnewline
 & KH80 L+P  & \textcolor{black}{$N(1675)\;5/2^{-}$} & $1654\pm2\pm0$ & $125\pm3\pm1$ & $23\pm1\pm0$ & %%@
$(-25\pm2\pm0)^{\circ}$\tabularnewline
 & KA84 L+P &  & $1656\pm1\pm0$ & $123\pm2\pm1$ & $23\pm1\pm0$ & $(-23\pm1\pm1)^{\circ}$\tabularnewline
\cline{2-7} 
 & RPP  &  & $\mathbf{2100\pm60}$ & $\mathbf{360\pm80}$ & $\mathbf{20\pm10}$ & $\mathbf{(-90\pm50)^{\circ}}$\tabularnewline
 & RPP H93 &  & $-$ & $-$ & $-$ & $-$\tabularnewline
 & KH80 L+P  & \textcolor{black}{$N(2060)^*\;5/2^{-}$} & $2119\pm11\pm1$ & $370\pm20\pm5$ & $19\pm1\pm1$ & %%@
$(-94\pm5\pm1)^{\circ}$\tabularnewline
 & KA84 L+P &  & $2134\pm9\pm5$ & $352\pm18\pm7$ & $18\pm1\pm1$ & $(-80\pm4\pm2)^{\circ}$\tabularnewline
\hline \hline 
\multirow{8}{*}{$F_{15}$} & RPP  &  & $\mathbf{1665-1680}$ & $\mathbf{110-135}$ & $\mathbf{40\pm5}$ & %%@
$\mathbf{(-10\pm10){}^{\circ}}$\tabularnewline
 & RPP H93 &  & $\mathbf{1673}$ & $\mathbf{135}$ & $\mathbf{44}$ & $\mathbf{-17^{\circ}}$\tabularnewline
 & KH80 L+P  & $N(1680)\:5/2^{+}$ & $1674\pm2\pm1$ & $129\pm3\pm1$ & $44\pm1\pm1$ & $(-16\pm1\pm1)^{\circ}$\tabularnewline
 & KA84 L+P &  & $1672\pm2\pm1$ & $132\pm4\pm1$ & $45\pm2\pm1$ & $(-16\pm2\pm1)^{\circ}$\tabularnewline
\cline{2-7} 
 & RPP  && $\begin{array}{c}
\mathbf{2030\pm110}\\
\mathbf{or\;1779}
\end{array}$ & $\begin{array}{c}
\mathbf{480\pm100}\\
\mathbf{or\:248}
\end{array}$ & $\mathbf{10-115}$ & $\mathbf{(-100\pm40)^{\circ}}$\tabularnewline
 & RPP H93 &  & $-$ & $-$ & $-$ & $-$\tabularnewline
 & KH80 L+P  &  $N(2000)^*\:5/2^{+}$   & $1834\pm19\pm6$ & $122\pm34\pm7$ & $4\pm1\pm1$ & $(-39\pm18\pm9)^{\circ}$\tabularnewline
 & KA84 L+P &  & $1838\pm20\pm25$ & $182\pm40\pm25$ & $5\pm2\pm1$ & $(-39\pm20\pm27)^{\circ}$\tabularnewline
\hline \hline
\multirow{4}{*}{$G_{17}$} & RPP  &  & $\mathbf{2050-2100}$ & $\mathbf{400-520}$ & $\mathbf{30-72}$ & $\mathbf{(-30\: %%@
to\:30)^{\circ}}$\tabularnewline
 & RPP H93 &  & $\mathbf{2042}$ & $\mathbf{482}$ & $\mathbf{45}$ & $-$\tabularnewline
 & KH80 L+P  & $N(2190)\:7/2^{+}$ & $2079\pm4\pm9$ & $509\pm7\pm16$ & $54\pm1\pm3$ & $(-18\pm1\pm3)^{\circ}$\tabularnewline
 & KA84 L+P &  & $2065\pm3\pm11$ & $526\pm7\pm2$ & $59\pm1\pm1$ & $(-22\pm1\pm5)^{\circ}$\tabularnewline
\hline \hline
\multirow{4}{*}{$G_{19}$} & RPP  & & \textbf{$\mathbf{2150-2250}$} & \textbf{$\mathbf{350-550}$} & \textbf{$\mathbf{20-30}$} & %%@
$\mathbf{(-50\pm30)^{\circ}}$\tabularnewline
 & RPP H93 &  & \textbf{$\mathbf{2187}$} & \textbf{$\mathbf{388}$} & \textbf{$\mathbf{21}$} & -\tabularnewline
 & KH80 L+P  &  \textbf{$N(2250)\:9/2^{-}$} & $2157\pm3\pm14$ & $412\pm7\pm44$ & $24\pm1\pm5$ & $(-62\pm1\pm11)^{\circ}$\tabularnewline
 & KA84 L+P &  & $2187\pm3\pm4$ & $396\pm6\pm19$ & $22\pm1\pm2$ & $(-41\pm1\pm3)^{\circ}$\tabularnewline
\hline \hline
\multirow{4}{*}{$H_{19}$} & RPP  & & $\mathbf{2130-2200}$ & $\mathbf{400-560}$ & $\mathbf{33-60}$ & %%@
$\mathbf{(-45\pm25)^{\circ}}$\tabularnewline
 & RPP H93 &  & $\mathbf{2135}$ & $\mathbf{400}$ & $\mathbf{40}$ & $\mathbf{-50^{\circ}}$\tabularnewline
 & KH80 L+P  &  $N(2220)\:9/2^{+}$ & $2127\pm3\pm24$ & $380\pm7\pm22$ & \textcolor{red}{${\color{black}38\pm1\pm5}$} & %%@
\textcolor{black}{$(-52\pm1\pm14)^{\circ}$}\tabularnewline
 & KA84 L+P &  & $2139\pm3\pm3$ & $390\pm6\pm1$ & $41\pm1\pm1$ & $(-48\pm1\pm1)^{\circ}$\tabularnewline
\hline \hline
\end{tabular}
\end{table*}


% Preview source code for paragraph 4

\begin{table*}[p]
{\caption{{\footnotesize \label{tab:pole3} Pole positions in MeV and residues
of partial waves as moduli in MeV and phases in degrees for lowest $I=3/2$ partial waves. The results
from L+P expansion are given for Karlsruhe-Helsinki 80 (KH80) and
Karlsruhe 84 (KA84) analyses. Resonances marked with a star indicate resonances which can be explained by $\rho$N complex branch point. %%@
{\red RPP denotes the range of pole parameters given by Ref.~\cite{PDG}, and RPP H93 denotes the values of pole parameters named %%@
HOEHLER 93 in RPP, and taken over from RPP and Table 1 of Ref. \cite{Hoehler93}.} \\ }}
}
\begin{tabular}{cc|ccccc}
\hline 
PW  & Source  & \textcolor{black}{Resonance}  & \textcolor{black}{$\;\mbox{Re}\, W_{p}\;$}  & \textcolor{black}{$\;-2\mbox{Im}\, %%@
W_{p}\;$}  & \textcolor{black}{$|\mbox{residue}|$}  & \textcolor{black}{$\theta$} \tabularnewline
\hline 
\hline 
\multirow{8}{*}{ $S_{31}$ } & RPP  & & $\mathbf{1590-1610}$ & $\mathbf{120-140}$ & $\mathbf{13-20}$ & %%@
$\mathbf{(-110\pm20)^{\circ}}$\tabularnewline
 & RPP H93  &  & $\mathbf{1608}$ & $\mathbf{116}$ & $\mathbf{19}$ & $\mathbf{-95^{\circ}}$\tabularnewline
 & \multicolumn{1}{c|}{KH80 L+P } &  $\Delta(1620)\:1/2^{-}$  & $1603\pm7\pm2$ & $114\pm12\pm4$ & $17\pm2\pm1$ & %%@
$(-106\pm10\pm4)^{\circ}$\tabularnewline
 & KA84 L+P & & $1605\pm5\pm2$ & $108\pm9\pm1$ & $16\pm0\pm1$ & $(-103\pm6\pm3)^{\circ}$\tabularnewline
\cline{2-7} 
 & RPP  &  & ${\begin{array}{c}
\mathbf{1820-1910}\\
\mathbf{or}\:\mathbf{1780}
\end{array}}$ & $\mathbf{130-345}$ & $\mathbf{10\pm3}$ & $\begin{array}{c}
\mathbf{(-125\pm20)^{\circ}}\\
\mathbf{or\:(20\pm40)^{\circ}}
\end{array}$\tabularnewline
 & RPP H93 &  & $\mathbf{1780}$ & $-$ & $-$ & $-$\tabularnewline
 & KH80 L+P  & $\Delta(1900)^*\:1/2^{-}$ & $1865\pm35\pm19$ & $187\pm50\pm19$ & $11\pm4\pm2$ & $(20\pm27\pm19)^{\circ}$\tabularnewline
 & KA84 L+P &  & $1867\pm22\pm9$ & $191\pm23\pm7$ & $12\pm0\pm2$ & $(22\pm11\pm8)^{\circ}$\tabularnewline
\hline \hline
\multirow{4}{*}{$P_{31}$ } & RPP  & & $\mathbf{1830-1880}$ & $\mathbf{200-500}$ & $\mathbf{16-45}$ & $\mathbf{-}$\tabularnewline
 & RPP H93 &  & $\mathbf{1874}$ & $\mathbf{283}$ & $\mathbf{38}$ & $-$\tabularnewline
 & KH80 L+P  & $\Delta(1910)\:1/2^{+}$   & $1896\pm11\pm0$ & $302\pm22\pm0$ & $29\pm2\pm0$ & $(-83\pm4\pm1)^{\circ}$\tabularnewline
 & KA84 L+P &  & $1880\pm19\pm11$ & $325\pm37\pm16$ & $30\pm4\pm1$ & $(-97\pm7\pm9)^{\circ}$\tabularnewline
\hline \hline

\multirow{12}{*}{$P_{33}$} & RPP  &  & $\mathbf{1209-1211}$ & $\mathbf{98-102}$ & $\mathbf{50\pm3}$ & %%@
$(-\mathbf{46\pm2)^{\circ}}$\tabularnewline
 & RPP H93 &  & $\mathbf{1209}$ & $\mathbf{100}$ & $\mathbf{50}$ & $-\mathbf{48^{\circ}}$\tabularnewline
 & KH80 L+P  & $\Delta(1232)\:3/2^{+}$ & $1211\pm1\pm1$ & $98\pm2\pm1$ & $50\pm1\pm1$ & $(-46\pm1\pm1)^{\circ}$\tabularnewline
 & KA84 L+P &  & $1210\pm1\pm1$ & $100\pm1\pm1$ & $51\pm1\pm1$ & $(-46\pm1\pm1)^{\circ}$\tabularnewline
\cline{2-7} 
 & RPP  &  & $\mathbf{1460-1560}$ & $\mathbf{200-350}$ & $\mathbf{5-44}$ & $\mathbf{-}$\tabularnewline
 & RPP H93  &  & $\mathbf{1550}$ & $-$ & $-$ & $-$\tabularnewline
 & KH80 L+P  & $\Delta(1600)\:3/2^{+}$ & $1469\pm10\pm5$ & $314\pm18\pm8$ & $38\pm2\pm2$ & $(173\pm5\pm5)^{\circ}$\tabularnewline
 & KA84 L+P &  & $1489\pm9\pm2$ & $289\pm17\pm6$ & $31\pm3\pm2$ & $(-174\pm5\pm3)^{\circ}$\tabularnewline
\cline{2-7} 
 & RPP  &  & $\mathbf{1850-1950}$ & $\mathbf{200-400}$ & $\mathbf{11-28}$ & $\mathbf{\mathbf{\begin{array}{c}
(-130\pm30)^{\circ}\\
(-45\pm30)^{\circ})
\end{array}}}$\tabularnewline
 & RPP H93 &  & $\mathbf{1900}$ & $-$ & $-$ & $-$\tabularnewline
 & KH80 L+P  & $\Delta(1920)^*\:3/2^{+}$ & $1906\pm10\pm2$ & $310\pm20\pm11$ & $26\pm3\pm2$ & $-(130\pm5\pm3)^{\circ}$\tabularnewline
 & KA84 L+P &  & $1923\pm9\pm2$ & $347\pm18\pm13$ & $31\pm2\pm2$ & $-(116\pm5\pm1)^{\circ}$\tabularnewline
\hline \hline
\multirow{8}{*}{$D_{33}$} & RPP  & & $\mathbf{1620-1680}$ & $\mathbf{160-300}$ & $\mathbf{10-50}$ & $\mathbf{(-45\: %%@
to\:12)^{\circ}}$\tabularnewline
 & RPP H93 &  & $\mathbf{1651}$ & $\mathbf{159}$ & $\mathbf{10}$ & -\tabularnewline
 & KH80 L+P  &  \textcolor{black}{$\Delta(1700)\;3/2^{-}$} & $1643\pm6\pm3$ & $217\pm10\pm8$ & $13\pm1\pm1$ & %%@
$(-30\pm4\pm3)^{\circ}$\tabularnewline
 & KA84 L+P &  & $1616\pm3\pm2$ & $280\pm6\pm3$ & $21\pm1\pm1$ & $(-58\pm2\pm2)^{\circ}$\tabularnewline
\cline{2-7} 
 & RPP  & & $\mathbf{1900-2080}$ & $\mathbf{190-400}$ & $\mathbf{1-8}$ & $\mathbf{(135\pm45)}^{\circ}$\tabularnewline
 & RPP H93 &  & $-$ & $-$ & $-$ & $-$\tabularnewline
 & KH80 L+P  & $\Delta(1940)^*\:3/2^{-}$  & $1878\pm11\pm5.5$ & $212\pm21\pm6$ & $9\pm1\pm1$ & $(140\pm7\pm7)^{\circ}$\tabularnewline
 & KA84 L+P &  & $1884\pm7\pm2$ & $303\pm13\pm8$ & $18\pm1\pm1$ & $(158\pm3\pm1)^{\circ}$\tabularnewline
\hline \hline
\multirow{4}{*}{$D_{35}$} & RPP  & & $\mathbf{1840-1960}$ & $\mathbf{175-360}$ & $\mathbf{7-30}$ & %%@
$\mathbf{(-20\pm40)^{\circ}}$\tabularnewline
 & RPP H93 &  & $\mathbf{1850}$ & $\mathbf{180}$ & $\mathbf{20}$ & $-$\tabularnewline
 & KH80 L+P  & \textcolor{black}{$\Delta(1930)\;5/2^{-}$}  & $1848\pm9\pm19$ & $321\pm17\pm7$ & $9\pm1\pm1$ & %%@
$(-37\pm3\pm7)^{\circ}$\tabularnewline
 & KA84 L+P &  & $1844\pm8\pm28$ & $334\pm17\pm9$ & $10\pm1\pm1$ & $(-40\pm3\pm9)^{\circ}$\tabularnewline
\hline \hline
\end{tabular}
\end{table*}

\begin{table*}[ht]
{\caption{{\footnotesize \label{tab:pole4} Pole positions in MeV and residues
of partial waves as moduli in MeV and phases in degrees for higher $I=3/2$ partial waves. The results
from L+P expansion are given for Karlsruhe-Helsinki 80 (KH80) and
Karlsruhe 84 (KA84) analyses. {\red RPP denotes the range of pole parameters given by Ref.~\cite{PDG}, and RPP H93 denotes the values %%@
of pole parameters named HOEHLER 93 in RPP, and taken over from RPP and Table 1 of Ref. \cite{Hoehler93}.} \\ }}
}
\begin{tabular}{cc|ccccc}
\hline 
PW  & Source  & \textcolor{black}{Resonance}  & \textcolor{black}{$\;\mbox{Re}\, W_{p}\;$}  & \textcolor{black}{$\;-2\mbox{Im}\, %%@
W_{p}\;$}  & \textcolor{black}{$|\mbox{residue}|$}  & \textcolor{black}{$\theta$} \tabularnewline
\hline 
\hline 
\multirow{8}{*}{$F_{35}$} & RPP  & & $\mathbf{1805-1835}$ & $\mathbf{265-300}$ & $\mathbf{25\pm10}$ & $\mathbf{(-50\pm20)^{\circ}}$ %%@
\tabularnewline
 & RPP H93 &  & $\mathbf{1829}$ & $\mathbf{303}$ & $\mathbf{25}$ & -\tabularnewline
 & KH80 L+P  &  $\Delta(1905)\:5/2^{+}$ & $1752\pm3\pm2$ & $346\pm6\pm2$ & $24\pm1\pm1$ & $-(114\pm1\pm2)^{\circ}$\tabularnewline
 & KA84 L+P &  & $1790\pm3\pm2$ & $293\pm6\pm6$ & $19\pm1\pm1$ & $-(77\pm2\pm2)^{\circ}$\tabularnewline
\cline{2-7} 
 & RPP  & & $\mathbf{2000}$ & $\mathbf{250-450}$ & $\mathbf{16\pm5}$ & $\mathbf{(150\pm90)^{\circ}}$\tabularnewline
 & RPP H93 &  & $-$ & $-$ & $-$ & $-$\tabularnewline
 & KH80 L+P  & $\Delta(2000)\:5/2^{+}$   & $1998\pm4\pm4$ & $404\pm10\pm4$ & $34\pm1\pm1$ & $(110\pm1\pm3)^{\circ}$\tabularnewline
 & KA84 L+P &  & $2035\pm6\pm6$ & $381\pm13\pm20$ & $23\pm1\pm3$ & $(132\pm2\pm5)^{\circ}$\tabularnewline
\hline \hline
\multirow{8}{*}{$F_{37}$} & RPP  &  & $\mathbf{1870-1890}$ & $\mathbf{220-260}$ & $\mathbf{47-61}$ & %%@
$\mathbf{(-33\pm12)^{\circ}}$\tabularnewline
 & RPP H93 &  & $\mathbf{1878}$ & $\mathbf{230}$ & $\mathbf{47}$ & $\mathbf{-32^{\circ}}$\tabularnewline
 & KH80 L+P  & $\Delta(1950)\:7/2^{+}$  & $1877\pm2\pm1$ & $223\pm4\pm1$ & $44\pm1\pm0$ & $-(39\pm1\pm1)^{\circ}$\tabularnewline
 & KA84 L+P &  & $1878\pm2\pm1$ & $246\pm4\pm3$ & $53\pm1\pm1$ & $-(36\pm1\pm1)^{\circ}$\tabularnewline
\cline{2-7} 
 & RPP  & & $\mathbf{2250-2350}$ & $\mathbf{160-360}$ & $\mathbf{12\pm6}$ & $\mathbf{(-90\pm60)^{\circ}}$\tabularnewline
 & RPP H93 &  & $-$ & $-$ & $-$ & $-$\tabularnewline
 & KH80 L+P  & $\Delta(2390)\:7/2^{+}$  & $2223\pm15\pm19$ & $431\pm26\pm7$ & $26\pm2\pm1$ & $(-160\pm5\pm11)^{\circ}$\tabularnewline
 & KA84 L+P &  & $2257\pm13\pm8$ & $472\pm25\pm20$ & $30\pm2\pm2$ & $(-131\pm4\pm3)^{\circ}$\tabularnewline
\hline \hline
\multirow{4}{*}{$H_{311}$} & RPP  &  & $\mathbf{2260-2400}$ & $\mathbf{350-750}$ & $\mathbf{12-39}$ & %%@
$\mathbf{-(30\pm40)^{\circ}}$\tabularnewline
 & RPP H93  &  & $\mathbf{2300}$ & $\mathbf{620}$ & $\mathbf{39}$ & $\mathbf{-60^{\circ}}$\tabularnewline
 & KH80 L+P  & $\Delta(2390)\:11/2^{+}$  & $2454\pm4\pm11$ & $462\pm8\pm50$ & $30\pm1\pm7$ & $(11\pm1\pm8)^{\circ}$\tabularnewline
 & KA84 L+P &  & $2301\pm3\pm4$ & $533\pm6\pm11$ & $31\pm1\pm1$ & $(-65\pm1\pm2)^{\circ}$\tabularnewline
\hline \hline \\ \\ \\ \\
\end{tabular}
\end{table*}

\subsection{Complex branch points} 
In Table \ref{tab:pole5} we give the parameters for some typical situations when fits with complex branch points achieve the similar %%@
quality as fits with the real ones (measured by the size of discrepancy variable $D_{dp}$, see Eq. \ref{def:D} ). 
The complex branch point is a mathematical implementation of the situation when the three-body final state contains a two body %%@
sub-channel accompanied by the third ``observer" particle. 
In this case we also allow for an extra resonance in the subchannel, but it is not located in the isobar intermediate state, but in the %%@
final state interaction. 
Both mechanisms (real and complex branch points) are indistinguishable in a single channel model. 
As was the case in  the J\"{u}lich model for $P_{11}$(1710), other channels (in the J\"{u}lich model K$\Lambda$ channel) are essential %%@
to distinguish between the two. 
Taking into account arbitrariness of 2- vs 3-body solutions, we list resonances which are quite well established by the %%@
Karlsruhe-Helsinki analysis, 
and those which are only possible depending on the ratio of two-body to three-body final state. 
We have tried to fit with complex branch points and more resonances;  
however, without knowing the branching fraction of two-body to three-body, the complex branch point takes over the whole flux, and %%@
eliminates the additional resonance altogether.

\begin{table*}[ht]
{\footnotesize \caption{{\footnotesize \label{tab:pole5} Pole positions in MeV and residues
of multipoles as moduli in mfm$\,\cdot\,$GeV and phases in degrees. $N_r$ is  number of resonance poles.
The results from L+P expansion are given for KH80  and KA84  solutions using $\rho N$ complex
branch point.}}
}
\begin{tabular}{cc|c|ccccccccc}
\hline \hline
Multipole  & Source  & $N_{r}$ & \textcolor{black}{Resonance}  & \textcolor{black}{$\;\mbox{Re}\, W_{p}\;$ }  & %%@
\textcolor{black}{$\;-2\mbox{Im}\, W_{p}\;$ }  & \textcolor{black}{$|\mbox{residue}|$ }  & \textcolor{black}{$\theta$ }  & $x_{P}$ & %%@
$x_{Q}$ & $x_{R}$ & $D_{dp}$\tabularnewline
\hline 
\hline 
$P_{11}$  & KH80 L+P  & 1 & $N(1440)\:1/2^{+}$  & $1368$  & $167$  & $45$  & $-81^{\circ}$ & $369$ & $1077^{\pi N}$ & $(1708-\i %%@
70)^{\rho N}$ & $0.325$\tabularnewline
 & KA84 L+P & 1 &  & $1372$  & $170$  & $43$  & $-78^{\circ}$ & $277$ & $1077^{\pi N}$ & $(1708-\i 70)^{\rho N}$ & %%@
$0.354$\tabularnewline
\hline \hline
$P_{13}$ & KH80 L+P  & 1 & $N(1720)\:/3/2^{+}$ & $1656$ & $175$ & $8$ & $-151^{\circ}$ & $385$ & $1077^{\pi N}$ & $(1708-\i 70)^{\rho %%@
N}$ & $0.119$\tabularnewline
 & KA84 L+P & 1 &  & $1676$ & $169$ & $10$ & $-105^{\circ}$ & $197$ & $1077^{\pi N}$ & $(1708-\i 70)^{\rho N}$ & $0.026$\tabularnewline
\hline \hline
$D_{13}$ & KH80 L+P  & 1 & \textcolor{black}{$N(1720)\;3/2^{-}$}  & $1506$  & $119$  & $34$  & $-15{}^{\circ}$ & $784$ & $1077^{\pi N}$ %%@
& $(1708-\i 70)^{\rho N}$ & $0.154$\tabularnewline
 & KA84 L+P & 1 &  & $1507$  & $114$  & $32$  & $-14{}^{\circ}$ & $756$ & $1077^{\pi N}$ & $(1708-\i 70)^{\rho N}$ & %%@
$0.161$\tabularnewline
\hline \hline
$D_{15}$ & KH80 L+P  & 1 & $N(1675)\:5/2^{-}$ & $1650$ & $88$ & $9$ & $-24^{\circ}$ & $454$ & $1077^{\pi N}$ & $(1708-\i 70)^{\rho N}$ %%@
& $0.471$\tabularnewline
 & KA84 L+P & 1 &  & $1656$ & $137$ & $29$ & $-30^{\circ}$ & $-644$ & $1077^{\pi N}$ & $(1708-\i 70)^{\rho N}$ & $0.058$\tabularnewline
\hline \hline
$F_{15}$ & KH80 L+P  & 1 & \textcolor{black}{$N(1680)\;5/2^{+}$}  & $1671$  & $142$  & $49$  & $-22{}^{\circ}$ & $176$ & $1077^{\pi N}$ %%@
& $(1708-\i 70)^{\rho N}$ & $0.071$\tabularnewline
 & KA84 L+P & 1 &  & $1674$  & $153$  & $46$  & $-23{}^{\circ}$ & $484$ & $1077^{\pi N}$ & $(1708-\i 70)^{\rho N}$ & %%@
$0.031$\tabularnewline
\hline \hline
$S_{31}$ & KH80 L+P  & 1 & $\Delta(1620)\:1/2^{-}$ & $1605$ & $139$ & $26$ & $-109^{\circ}$ & $45$ & $1077^{\pi N}$ & $(1708-\i %%@
70)^{\rho N}$ & $0.021$\tabularnewline
 & KA84 L+P & 1 &  & $1605$ & $128$ & $21$ & $-107^{\circ}$ & $-2446$ & $1077^{\pi N}$ & $(1708-\i 70)^{\rho N}$ & %%@
$0.018$\tabularnewline
\hline \hline
$P_{31}$ & KH80 L+P  & 1 & $\Delta(1910)\:1/2^{+}$ & $1847$ & $257$ & $49$ & $-128^{\circ}$ & $739$ & $1077^{\pi N}$ & $(1708-\i %%@
70)^{\rho N}$ & $0.109$\tabularnewline
 & KA84 L+P & 1 &  & $1891$ & $398$ & $40$ & $-75^{\circ}$ & $-203$ & $1077^{\pi N}$ & $(1708-\i 70)^{\rho N}$ & $0.025$\tabularnewline
\cline{2-12}
 & KH80 L+P  & 0 & - & - & - & - & - & $-556$ & $1077^{\pi N}$ & $(1708-\i 70)^{\rho N}$ & $0.123$\tabularnewline
 & KA84 L+P & 0 & - & - &  & - &  & $404$ & $1077^{\pi N}$ & $(1708-\i 70)^{\rho N}$ & $0.040$\tabularnewline
\hline \hline
$P_{33}$ & KH80 L+P  & \multirow{2}{*}{2} & $\Delta(1232)\:3/2^{+}$ & $1210$ & $102$ & $53$ & $-47^{\circ}$ & \multirow{2}{*}{$656$} & %%@
\multirow{2}{*}{$1077^{\pi N}$} & \multirow{2}{*}{$(1708-\i 70)^{\rho N}$} & \multirow{2}{*}{$0.025$}\tabularnewline
 &  &  & $\Delta(1600\:3/2^{+}$ & $1537$ & $157$ & $10$ & $-105^{\circ}$ &  &  &  & \tabularnewline
\cline{2-12} 
 & KA84 L+P & \multirow{2}{*}{2} & $\Delta(1232)\:3/2^{+}$ & $1210$ & $102$ & $53$ & $-47^{\circ}$ & \multirow{2}{*}{$-403$} & %%@
\multirow{2}{*}{$1077^{\pi N}$} & \multirow{2}{*}{$(1708-\i 70)^{\rho N}$} & \multirow{2}{*}{$0.034$}\tabularnewline
 &  &  & $\Delta(1600\:3/2^{+}$ & $1545$ & $155$ & $10$ & $-95^{\circ}$ &  &  &  & \tabularnewline
\hline \hline
$D_{33}$ & KH80 L+P  & \multicolumn{1}{c|}{1} & $\Delta(1700)\:3/2^{+}$ & $1663$ & $180$ & $12$ & $15$ & \multicolumn{1}{c}{$53$} & %%@
\multicolumn{1}{c}{$1077^{\pi N}$} & \multicolumn{1}{c}{$(1708-\i 70)^{\rho N}$} & \multicolumn{1}{c}{$0.161$}\tabularnewline
 & KA84 L+P & 1 &  & $1574$ & $373$ & $29$ & $-111^{\circ}$ & $69$ & $1077^{\pi N}$ & $(1708-\i 70)^{\rho N}$ & $0.034$\tabularnewline
\hline \hline
$D_{35}$ & KH80 L+P  & 1 & $\Delta(1930)\:5/2^{-}$ & $1813$ & $242$ & $8$ & $-72^{\circ}$ & $-302$ & $1077^{\pi N}$ & $(1708-\i %%@
70)^{\rho N}$ & $0.498$\tabularnewline
 & KA84 L+P & 1 &  & $1889$ & $258$ & $16$ & $-49^{\circ}$ & $-2398$ & $1077^{\pi N}$ & $(1708-\i 70)^{\rho N}$ & %%@
$0.069$\tabularnewline
\cline{2-12}
 & KH80 L+P  & 0 & - & - & - & - & - & $887$ & $1077^{\pi N}$ & $(1708-\i 70)^{\rho N}$ & $0.303$\tabularnewline
 & KA84 L+P & 0 & - & - & - & - & - & $24$ & $1077^{\pi N}$ & $(1708-\i 70)^{\rho N}$ & $0.102$\tabularnewline
\hline \hline
$F_{35}$ & KH80 L+P  & \multirow{2}{*}{2} & $\Delta(1905)\:5/2^{+}$ & $1782$ & $243$ & $17$ & $-162^{\circ}$ & \multirow{2}{*}{$899$} & %%@
\multirow{2}{*}{$1077^{\pi N}$} & \multirow{2}{*}{$(1708-\i 70)^{\rho N}$} & \multirow{2}{*}{$0.254$}\tabularnewline
 &  &  & $\Delta(2000)\:5/2^{+}$ & $2027$ & $449$ & $39$ & $137^{\circ}$ &  &  &  & \tabularnewline
\cline{2-12}
 & KA84 L+P & \multirow{2}{*}{2} & $\Delta(1905)\:5/2^{+}$ & $1790$ & $314$ & $22$ & $-76^{\circ}$ & \multirow{2}{*}{$-240$} & %%@
\multirow{2}{*}{$1077^{\pi N}$} & \multirow{2}{*}{$(1708-\i 70)^{\rho N}$} & \multirow{2}{*}{$0.045$}\tabularnewline
 &  &  & $\Delta(2000)\:5/2^{+}$ & $2035$ & $408$ & $27$ & $135^{\circ}$ &  &  &  & \tabularnewline
\hline \hline
$F_{37}$ & KH80 L+P  & \multirow{2}{*}{2} & $\Delta(1950)\:7/2^{+}$ & $1893$ & $275$ & $65$ & $-15^{\circ}$ & \multirow{2}{*}{$802$} & %%@
\multirow{2}{*}{$1077^{\pi N}$} & \multirow{2}{*}{$(1708-\i 70)^{\rho N}$} & \multirow{2}{*}{$0.204$}\tabularnewline
 &  &  & $\Delta(2390)\:7/2^{+}$ & $2419$ & $323$ & $16$ & $-27^{\circ}$ &  &  &  & \tabularnewline
\cline{2-12}
 & KA84 L+P & \multirow{2}{*}{2} & $\Delta(1950)\:7/2^{+}$ & $1882$ & $252$ & $54$ & $-31^{\circ}$ & \multirow{2}{*}{$520$} & %%@
\multirow{2}{*}{$1077^{\pi N}$} & \multirow{2}{*}{$(1708-\i 70)^{\rho N}$} & \multirow{2}{*}{$0.021$}\tabularnewline
 &  &  & $\Delta(2390)\:7/2^{+}$ & $2311$ & $469$ & $31$ & $-100^{\circ}$ &  &  &  & \tabularnewline
\hline \hline
\end{tabular}
\end{table*}

\clearpage
\section{Discussion and Conclusions}
\noindent
{\red
Using the L+P method we obtain almost perfect fits to all KH80 and KA84 partial waves. 
This is visible in very low discrepancy parameters $D_{dp}$ given in Tables~\ref{tab:paramKH80} and \ref{tab:paramKA84}, 
and in excellent visual agreement of fitting  curves and input data in Figs. \ref{Fig1} - \ref{Fig4}. 
Agreement is somewhat poorer for KH80 H$_{311}$, G$_{17}$ and G$_{19}$ partial waves. 
For the first one, the discrepancy parameter is of the order of 2-3, while for the latter two it is of order  1. 
For all other partial waves it is significantly below one. 
However, these results are consistent with the graphs in Fig. \ref{Fig1} - \ref{Fig4}. 
A closer look at the KH80 H$_{3,11}$ partial wave in  Fig.~\ref{Fig4}, one can detect somewhat poorer agreement of the real part of the %%@
fitted curve with data near 1700 and 2100 MeV, and for imaginary parts near 1900 MeV. 
Similar discrepancies can be found for G$_{17}$  and G$_{19}$ partial waves if Fig.~\ref{Fig2} is closely inspected. 
Such exceptions are not present for the KA84 solution. 
All partial waves for KA84 are fitted with discrepancy parameters significantly below one. 
This supports the statement given in \cite{Hoehler93} that the KA84 solution is obtained by further smoothing of the KH80 solution, 
and the L+P method can fit KA84 slightly better than KH80 due to additional smoothing. }
\\ \\ \noindent
We confirm the values of all the pole positions of the Karlsruhe-Helsinki solutions  given in the RPP  using the speed plot method of %%@
Ref~\cite{Hoehler93} for the KA84 solution  with better precision and confidence, 
giving corresponding solutions for the same resonances of the KH80 solution,  and a number of new poles which all agree with results %%@
quoted in the RPP.
\\ \\ \noindent
The new resonances, in the RPP but not established by the SP method are: S$_{11}$ $N(1895)1/2-$,  P$_{11}$ $N(2100)1/2+$,  P$_{13}$ %%@
$N(1900)3/2+$,  D$_{13}$ $N(1875)3/2-$,  F$_{15}$ $N(2000)5/2+$,  D$_{33}$ $\Delta(1940)3/2-$,  F$_{35}$ $\Delta(2000)5/2+$ and  %%@
F$_{37}$ $\Delta(2390)7/2+$.
\\ \\ \noindent
We confirm that visual shapes of KH80 and KA84 solutions and numerical values of pole positions are very similar, and in practice %%@
either solution can be used.
Masses (real parts) of KH80 and KA84 poles are within error bars; however some partial waves (the first D$_{33}$: $\Delta(1700) $3/2- %%@
and first F$_{35}$ $\Delta(1905)$ 5/2+) show slightly more  than one standard deviation discrepancy when the widths (imaginary parts) %%@
of poles are compared. 
Others are within one standard deviation.
\\ \\ \noindent
We establish that any single-channel model based solely on one channel of input data (in our case Karlsruhe-Helsinki PWA), is unable to %%@
distinguish between alternative two-body and three-body final state solutions. 
The L+P model can produce equivalent solutions for two-body and three-body final states; 
without new data, the two-body to three-body branching fraction remains undetermined. 
In Tables \ref{tab:pole1}, \ref{tab:pole2}, \ref{tab:pole3} and \ref{tab:pole4} we denote with asterisks solutions which have the same %%@
discrepancy ratio, but realized through different physics formalisms: 
two-body final state given with real branch point or three-body final state given by complex branch point.  
All these solutions are indistinguishable within single-channel models.
\\ \\ \noindent
It is very interesting to observe that we have even more ambiguity in the L+P method. 
There are partial waves in which the L+P method gives equivalent solutions in three-body formalism with one resonance or without any %%@
resonances at all. 
These are: P$_{31}$ and D$_{35}$ partial waves (see Table~\ref{tab:pole5}).
\\ \\ \noindent
What is important is that the dominant resonances in two-body or three-body formalism have identical parameters; i.e. the  single %%@
channel formalism \emph{without ambiguity} establishes the existence of resonances without an asterisk. 
\\ \\ \noindent
 The three-body formalism using  complex branch points raises some doubt about higher order resonances, and requires measurement of new %%@
data for inelastic channels.
 Only firm experimental numbers on inelastic 2-body $\rightarrow$ 2-body or higher energy 2-body $\rightarrow$ 3-body data can resolve %%@
the ambiguity between solutions given by single-channel analysis.   
We endorse strongly any new proposal which plans to measure inelastic $\pi N \rightarrow XY$ channels, e.g. \cite{Hicks2013}.

\section*{acknowledgment}
Many thanks go to David Bugg who carefully read the manuscript, significantly improved English fluency, and helped this paper to attain %%@
its final form. 
This work was suported in part by Tuzla Canton, Bosnia and Herzegovina, Ministry of Education, Science, Culture and Sport grant %%@
10/1-14-020637-1/13.
\clearpage
\section*{appendix}
\begin{table*}[!hb]
\caption{\label{tab:paramKH80} Parameters from L+P expansion are given for KH80 solution. 
$N_r$ is number of resonance poles, $x_P, x_Q, x_R$ are branch points in \vspace*{1.cm} MeV.}
\begin{tabular}{c|crclc|ccrclc}
\hline \hline
\multicolumn{12}{c}{Source KH80}\tabularnewline
\hline \hline
PW & $N_{r}$ & $x_{P}$ & $x_{Q}$ & $x_{R}$ & $D_{dp}$ & PW & $N_{r}$ & $x_{P}$ & $x_{Q}$ & $x_{R}$ & $D_{dp}$\tabularnewline
\hline \hline
\multirow{4}{*}{$S_{11}$} & $3$ & $-18216$ & $1077^{\pi N}$ & $1215^{\pi\pi N}$ & $0.131$ & \multirow{4}{*}{$S_{31}$} & $2$ & $-1123$ & %%@
$1077^{\pi N}$ & $1215^{\pi\pi N}$ & $0.036$\tabularnewline
 & $3$ & $779$ & $1077^{\pi N}$ & $1486^{\eta N}$ & $0.130$ &  & $2$ & $-1967$ & $1077^{\pi N}$ & $1370^{Real(\pi\Delta)}$ & %%@
$0.043$\tabularnewline
 & $3$ & $-2529$ & $1077^{\pi N}$ & $1491^{free}$ & $0.127$ &  & $2$ & $900$ & $1077^{\pi N}$ & $1708^{Real(\rho N)}$ & %%@
$0.041$\tabularnewline
 &  &  &  &  &  &  & 2 & $-1239$ & $1077^{\pi N}$ & $1702^{free}$ & $0.035$\tabularnewline
\hline \hline
\multirow{3}{*}{$P_{11}$} & $3$ & $-1135$ & $1077^{\pi N}$ & $1215^{\pi\pi N}$ & $0.408$ & \multirow{3}{*}{$P_{31}$} & $1$ & $314$ & %%@
$1077^{\pi N}$ & $1215^{\pi\pi N}$ & $0.0789$\tabularnewline
 & $3$ & $-1270$ & $1077^{\pi N}$ & $1370^{Real(\pi\Delta)}$ & $0.453$ &  & $1$ & $281$ & $1077^{\pi N}$ & $1210^{free}$ & %%@
$0.0786$\tabularnewline
 & $3$ & $-1988$ & $1077^{\pi N}$ & $1320^{free}$ & $0.474$ &  &  &  &  &  & \tabularnewline
\hline \hline
\multirow{3}{*}{$P_{13}$} & $2$ & $-28412$ & $1077^{\pi N}$ & $1215^{\pi\pi N}$ & $0.126$ & \multirow{3}{*}{$P_{33}$} & $3$ & $215$ & %%@
$1077^{\pi N}$ & $1215^{\pi\pi N}$ & $0.098$\tabularnewline
 & $2$ & $776$ & $1077^{\pi N}$ & $1370^{Real(\pi\Delta)}$ & $0.119$ &  & $3$ & $707$ & $1077^{\pi N}$ & $1370^{Real(\pi\Delta)}$ & %%@
$0.097$\tabularnewline
 & $2$ & $-617$ & $1077^{\pi N}$ & $1267$ & $0.118$ &  & $3$ & $898$ & $1077^{\pi N}$ & $1378^{free}$ & $0.076$\tabularnewline
\hline \hline 
 & $3$ & $-697$ & $1077^{\pi N}$ & $1215^{\pi\pi N}$ & $0.230$ & \multirow{3}{*}{$D_{33}$} & $2$ & $-3832$ & $1077^{\pi N}$ & %%@
$1215^{\pi\pi N}$ & $0.178$\tabularnewline
 & $3$ & $-4763$ & $1077^{\pi N}$ & $1370^{Real(\pi\Delta)}$ & $0.278$ &  & $2$ & $-3104$ & $1077^{\pi N}$ & $1370^{Real(\pi\Delta)}$ & %%@
$0.095$\tabularnewline
$D_{13}$ & $3$ & $-8507$ & $1077^{\pi N}$ & $1708^{Real(\rho N)}$ & $0.236$ &  & $2$ & $-14033$ & $1077^{\pi N}$ & $1362^{free}$ & %%@
$0.094$\tabularnewline
 & $3$ & $-2066$ & $1077^{\pi N}$ & $1107^{free}$ & $0.224$ &  &  &  &  &  & \tabularnewline
\hline \hline
\multirow{3}{*}{$D_{15}$} & $2$ & $407$ & $1077^{\pi N}$ & $1215^{\pi\pi N}$ & $0.536$ & \multirow{3}{*}{$D_{35}$} & $1$ & $315$ & %%@
$1077^{\pi N}$ & $1215^{\pi\pi N}$ & $0.576$\tabularnewline
 & $2$ & $223$ & $1077^{\pi N}$ & $1370^{Real(\pi\Delta)}$ & $0.525$ &  & $1$ & $331$ & $1077^{\pi N}$ & $1688^{K\Sigma}$ & %%@
$0.578$\tabularnewline
 & $2$ & $-5667$ & $1077^{\pi N}$ & $1511{}^{free}$ & $0.469$ &  & $1$ & $409$ & $1077^{\pi N}$ & $1211^{free}$ & %%@
$0.576$\tabularnewline
\hline \hline
\multirow{4}{*}{$F_{15}$} & $2$ & $239$ & $1077^{\pi N}$ & $1215^{\pi\pi N}$ & $0.136$ & \multirow{4}{*}{$F_{35}$} & $2$ & $7.8$ & %%@
$1077^{\pi N}$ & $1215^{\pi\pi N}$ & $0.343$\tabularnewline
 & $2$ & $43.9$ & $1077^{\pi N}$ & $1370^{Real(\pi\Delta)}$ & $0.124$ &  & $2$ & $-249$ & $1077^{\pi N}$ & $1708^{Real(\rho N)}$ & %%@
$0.344$\tabularnewline
 & $2$ & $-157$ & $1077^{\pi N}$ & $1708^{Real(\rho N)}$ & $0.108$ &  & $2$ & $98$ & $1077^{\pi N}$ & $1221^{free}$ & %%@
$0.330$\tabularnewline
 & $2$ & $-14.1$ & $1077^{\pi N}$ & $1673{}^{free}$ & $0.105$ &  &  &  &  &  & \tabularnewline
\hline \hline
\multirow{3}{*}{$G_{17}$} & $1$ & $-261$ & $1077^{\pi N}$ & $1215^{\pi\pi N}$ & $1.211$ & \multirow{3}{*}{$F_{37}$} & $2$ & $-324$ & %%@
$1077^{\pi N}$ & $1370^{Real(\pi\Delta)}$ & $0.376$\tabularnewline
 & $1$ & $298$ & $1077^{\pi N}$ & $1486^{\eta N}$ & $1.302$ &  & $2$ & $-439$ & $1077^{\pi N}$ & $1708^{Real(\rho N)}$ & %%@
$0.379$\tabularnewline
 & $1$ & $-148$ & $1077^{\pi N}$ & $1445^{free}$ & $1.164$ &  & $2$ & $-141$ & $1077^{\pi N}$ & $1463^{free}$ & $0.374$\tabularnewline
\hline \hline
\multirow{3}{*}{$G_{19}$} & $1$ & $-10490$ & $1077^{\pi N}$ & $1486^{\eta N}$ & $1.835$ & \multirow{3}{*}{$H_{311}$} & $1$ & $-35183$ & %%@
$1077^{\pi N}$ & $1215^{\pi\pi N}$ & $3.513$\tabularnewline
 & $1$ & $-838$ & $1077^{\pi N}$ & $1611^{K\Lambda}$ & $1.025$ &  & $1$ & $-1460$ & $1077^{\pi N}$ & $1688^{K\Sigma}$ & %%@
$3.009$\tabularnewline
 & $1$ & $-196$ & $1077^{\pi N}$ & $1713^{free}$ & $0.975$ &  & $1$ & $87.7$ & $1077^{\pi N}$ & $1489^{free}$ & $2.482$\tabularnewline
\hline \hline
\multirow{3}{*}{$H_{19}$} & $1$ & $-49$ & $1077^{\pi N}$ & $1486^{\eta N}$ & $0.315$ &  &  &  &  &  & \tabularnewline
 & $1$ & $-1093$ & $1077^{\pi N}$ & $1611^{K\Lambda}$ & $0.492$ &  &  &  &  &  & \tabularnewline
 & $1$ & $-1252$ & $1077^{\pi N}$ & $1709^{free}$ & $0.298$ &  &  &  &  &  & \tabularnewline
\cline{1-6}
\end{tabular}
\end{table*}

\begin{table*}[!ht]
\caption{\label{tab:paramKA84} Parameters from L+P expansion are given for KA84 solution. 
$N_r$ is number of resonance poles, $x_P, x_Q, x_R$ are branch points in  \vspace*{1.cm} MeV.}
\begin{tabular}{c|crclc|ccrclc}

\hline \hline
\multicolumn{12}{c}{Source KA84}\tabularnewline
\hline \hline
PW & $N_{r}$ & $x_{P}$ & $x_{Q}$ & $x_{R}$ & $D_{dp}$ & PW & $N_{r}$ & $x_{P}$ & $x_{Q}$ & $x_{R}$ & $D_{dp}$\tabularnewline
\hline \hline
\multirow{4}{*}{$S_{11}$} & $3$ & $822$ & $1077^{\pi N}$ & $1215^{\pi\pi N}$ & $0.159$ & \multirow{4}{*}{$S_{31}$} & $2$ & $-521$ & %%@
$1077^{\pi N}$ & $1215^{\pi\pi N}$ & $0.034$\tabularnewline
 & $3$ & $900$ & $1077^{\pi N}$ & $1486^{\eta N}$ & $0.105$ &  & $2$ & $663$ & $1077^{\pi N}$ & $1370^{Real(\pi\Delta)}$ & %%@
$0.039$\tabularnewline
 & $3$ & $900$ & $1077^{\pi N}$ & $1499^{free}$ & $0.096$ &  & $2$ & $196$ & $1077^{\pi N}$ & $1708^{Real(\rho N)}$ & %%@
$0.037$\tabularnewline
 &  &  &  &  &  &  &  & $-255$ & $1077^{\pi N}$ & $1217^{free}$ & $0.033$\tabularnewline
\hline \hline
\multirow{3}{*}{$P_{11}$} & $3$ & $287$ & $1077^{\pi N}$ & $1215^{\pi\pi N}$ & $0.459$ & \multirow{3}{*}{$P_{31}$} & $1$ & $-690$ & %%@
$1077^{\pi N}$ & $1215^{\pi\pi N}$ & $0.091$\tabularnewline
 & $3$ & $-7351$ & $1077^{\pi N}$ & $1370^{Real(\pi\Delta)}$ & $0.451$ &  & $1$ & $-658$ & $1077^{\pi N}$ & $1221^{free}$ & %%@
$0.088$\tabularnewline
 & $3$ & $-2082$ & $1077^{\pi N}$ & $1382{}^{free}$ & $0.377$ &  &  &  &  &  & \tabularnewline
\hline \hline
\multirow{3}{*}{$P_{13}$} & $2$ & $345$ & $1077^{\pi N}$ & $1215^{\pi\pi N}$ & $0.037$ & \multirow{3}{*}{$P_{33}$} & $3$ & $440$ & %%@
$1077^{\pi N}$ & $1215^{\pi\pi N}$ & $0.081$\tabularnewline
 & $2$ & $-957$ & $1077^{\pi N}$ & $1370^{Real(\pi\Delta)}$ & $0.038$ &  & $3$ & $576$ & $1077^{\pi N}$ & $1370^{Real(\pi\Delta)}$ & %%@
$0.088$\tabularnewline
 & $2$ & $543$ & $1077^{\pi N}$ & $1201^{free}$ & $0.036$ &  & $3$ & $-646$ & $1077^{\pi N}$ & $1469^{free}$ & $0.076$\tabularnewline
\hline \hline
\multirow{4}{*}{$D_{13}$} & $3$ & $-0.024$ & $1077^{\pi N}$ & $1215^{\pi\pi N}$ & $0.332$ & \multirow{3}{*}{$D_{33}$} & $2$ & $-696$ & %%@
$1077^{\pi N}$ & $1215^{\pi\pi N}$ & $0.066$\tabularnewline
 & $3$ & $-1567$ & $1077^{\pi N}$ & $1370^{Real(\pi\Delta)}$ & $0.331$ &  & $2$ & $-31769$ & $1077^{\pi N}$ & $1370^{Real(\pi\Delta)}$ %%@
& $0.063$\tabularnewline
 & $3$ & $-758$ & $1077^{\pi N}$ & $1708^{Real(\rho N)}$ & $0.270$ &  & $2$ & $-27693$ & $1077^{\pi N}$ & $1362^{free}$ & %%@
$0.061$\tabularnewline
 & $3$ & $-1449$ & $1077^{\pi N}$ & $1880{}^{free}$ & $0.249$ &  &  &  &  &  & \tabularnewline
\hline \hline
\multirow{3}{*}{$D_{15}$} & $2$ & $753$ & $1077^{\pi N}$ & $1215^{\pi\pi N}$ & $0.069$ & \multirow{3}{*}{$D_{35}$} & $1$ & $-3753$ & %%@
$1077^{\pi N}$ & $1215^{\pi\pi N}$ & $0.062$\tabularnewline
 & $2$ & $-4045$ & $1077^{\pi N}$ & $1370^{Real(\pi\Delta)}$ & $0.070$ &  & $1$ & $271$ & $1077^{\pi N}$ & $1688^{K\Sigma}$ & %%@
$0.063$\tabularnewline
 & $2$ & $-5667$ & $1077^{\pi N}$ & $1547{}^{free}$ & $0.057$ &  & $1$ & $409$ & $1077^{\pi N}$ & $1382^{free}$ & %%@
$0.060$\tabularnewline
\hline \hline
\multirow{4}{*}{$F_{15}$} & $2$ & $-139$ & $1077^{\pi N}$ & $1215^{\pi\pi N}$ & $0.084$ & \multirow{4}{*}{$F_{35}$} & $2$ & $-1063$ & %%@
$1077^{\pi N}$ & $1215^{\pi\pi N}$ & $0.045$\tabularnewline
 & $2$ & $-0.047$ & $1077^{\pi N}$ & $1370^{Real(\pi\Delta)}$ & $0.081$ &  & $2$ & $-3331$ & $1077^{\pi N}$ & $1708^{Real(\rho N)}$ & %%@
$0.057$\tabularnewline
 & $2$ & $-332$ & $1077^{\pi N}$ & $1708^{Real(\rho N)}$ & $0.052$ &  & $2$ & $-1384$ & $1077^{\pi N}$ & $1186^{free}$ & %%@
$0.044$\tabularnewline
 & $2$ & $546$ & $1077^{\pi N}$ & $1361{}^{free}$ & $0.027$ &  &  &  &  &  & \tabularnewline
\hline \hline
\multirow{3}{*}{$G_{17}$} & $1$ & $-50$ & $1077^{\pi N}$ & $1215^{\pi\pi N}$ & $0.354$ & \multirow{3}{*}{$F_{37}$} & $2$ & $-4046$ & %%@
$1077^{\pi N}$ & $1370^{Real(\pi\Delta)}$ & $0.039$\tabularnewline
 & $1$ & $-1513$ & $1077^{\pi N}$ & $1486^{\eta N}$ & $0.453$ &  & $2$ & $-3041$ & $1077^{\pi N}$ & $1708^{Real(\rho N)}$ & %%@
$0.039$\tabularnewline
 & $1$ & $250$ & $1077^{\pi N}$ & $1307^{free}$ & $0.351$ &  & $2$ & $-3167$ & $1077^{\pi N}$ & $1903^{free}$ & $0.027$\tabularnewline
\hline \hline
\multirow{3}{*}{$G_{19}$} & $1$ & $-1459$ & $1077^{\pi N}$ & $1486^{\eta N}$ & $0.345$ & \multirow{3}{*}{$H_{311}$} & $1$ & $-983$ & %%@
$1077^{\pi N}$ & $1215^{\pi\pi N}$ & $0.136$\tabularnewline
 & $1$ & $-2385$ & $1077^{\pi N}$ & $1611^{K\Lambda}$ & $0.556$ &  & $1$ & $-1099$ & $1077^{\pi N}$ & $1688^{K\Sigma}$ & %%@
$0.142$\tabularnewline
 & $1$ & $194$ & $1077^{\pi N}$ & $1406^{free}$ & $0.115$ &  & $1$ & $44$ & $1077^{\pi N}$ & $1462^{free}$ & $0.107$\tabularnewline
\hline \hline
\multirow{3}{*}{$H_{19}$} & $1$ & $-378$ & $1077^{\pi N}$ & $1486^{\eta N}$ & $0.027$ &  &  &  &  &  & \tabularnewline
 & $1$ & $433$ & $1077^{\pi N}$ & $1611^{K\Lambda}$ & $0.021$ &  &  &  &  &  & \tabularnewline
 & $1$ & $556$ & $1077^{\pi N}$ & $1715^{free}$ & $0.019$ &  &  &  &  &  & \tabularnewline
\cline{1-6} 
\end{tabular}
\end{table*}

\end{document}